\documentclass[jair,twoside,11pt,theapa]{article}
\usepackage{jair, theapa, rawfonts}

\jairheading{69}{2020}{807-845}{05/2020}{11/2020}
\ShortHeadings{Artificial Intelligence applications against COVID-19}
{Bullock, Luccioni, Hoffmann Pham, Lam, \& Luengo-Oroz}
\firstpageno{807}

\usepackage{graphicx}
\usepackage{navigator}
\newcommand{\preprint}{$^\dagger$}

\begin{document}

\title{Mapping the Landscape of Artificial Intelligence\\ Applications against COVID-19}

\author{\name Joseph Bullock \email joseph@unglobalpulse.org \\
        \addr United Nations Global Pulse,
        New York, NY, USA \\
        Institute for Data Science,
        Durham University, United Kingdom
\AND
       \name Alexandra Luccioni \email sasha.luccioni@mila.quebec  \\
        \addr  Mila Qu\'{e}bec Artificial Intelligence Institute \\
  Universit\'{e} de Montr\'{e}al,
  Montr\'{e}al, Canada
\AND
       \name Katherine Hoffmann Pham \email  katherine@unglobalpulse.org \\
        \addr 
        United Nations Global Pulse, New York, NY, USA \\
        NYU Stern School of Business, New York, NY, USA 
\AND
       \name Cynthia Sin Nga Lam \email  cynthia@unglobalpulse.org \\
        \addr United Nations Global Pulse,
        New York, NY, USA \\
        Global Coordination Mechanism on NCDs, \\ World Health Organization, Geneva, Switzerland
\AND
       \name Miguel Luengo-Oroz \email miguel@unglobalpulse.org \\
        \addr United Nations Global Pulse,
        New York, NY, USA
}

\maketitle

\begin{abstract} 
COVID-19, the disease caused by the SARS-CoV-2 virus, has been declared a pandemic by the World Health Organization, which has reported over 18 million confirmed cases as of August 5, 2020. In this review, we present an overview of recent studies using Machine Learning and, more broadly, Artificial Intelligence, to tackle many aspects of the COVID-19 crisis. We have identified applications that address challenges posed by COVID-19 at different scales, including: \textit{molecular}, by identifying new or existing drugs for treatment; \textit{clinical}, by supporting diagnosis and evaluating prognosis based on medical imaging and non-invasive measures; and \textit{societal}, by tracking both the epidemic and the accompanying infodemic using multiple data sources. We also review datasets, tools, and resources needed to facilitate Artificial Intelligence research, and discuss strategic considerations related to the operational implementation of multidisciplinary partnerships and open science. We highlight the need for international cooperation to maximize the potential of AI in this and future pandemics. 
\end{abstract}
~\\

\section{Introduction}

With the continued growth of the COVID-19 pandemic, researchers worldwide are working to better understand and suppress its spread. Key areas of research include studying COVID-19 transmission, facilitating its detection, developing possible vaccines and treatments, and understanding the socio-economic impacts of the pandemic. In this article, we discuss how Artificial Intelligence (AI) can contribute to these goals by enhancing ongoing research efforts, improving the efficiency and speed of existing approaches, and proposing original lines of research.

\begin{figure}
\centering
\includegraphics[width=\linewidth]{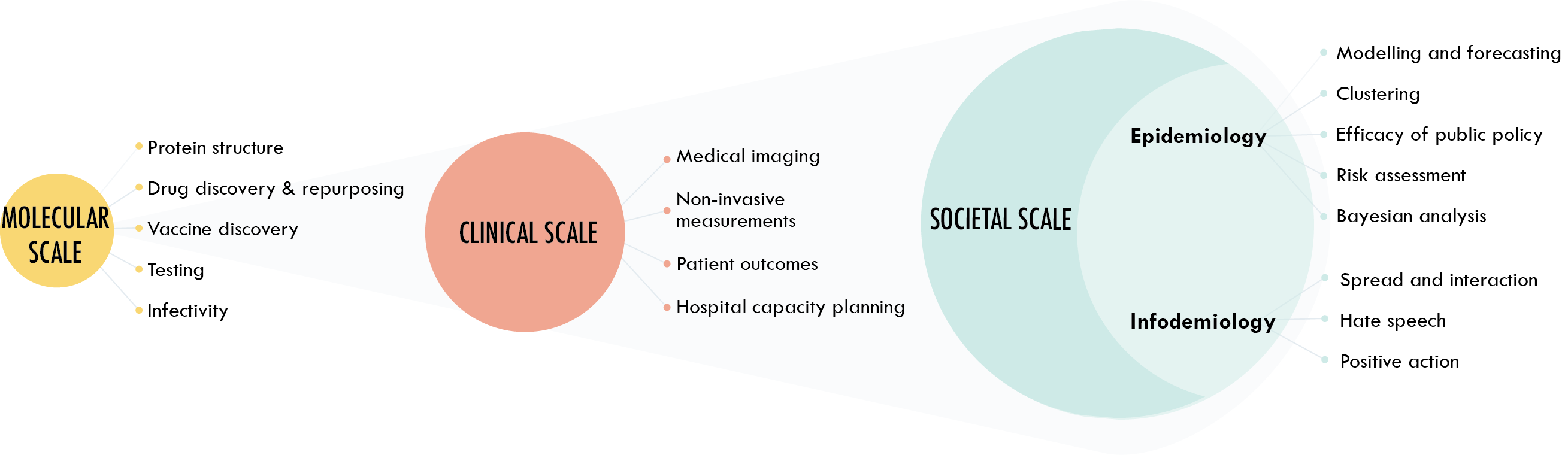}
\caption{AI applications for the COVID-19 response organized at three levels: the molecular scale, the clinical scale, and the societal scale. }\label{fig:scales}
\end{figure}

We have conducted an extensive review of the rapidly emerging literature and identified specific applications of AI at three different scales:\footnote{The election of this multi-scale categorization was inspired by the digital \textit{in-toto} reconstruction of living organisms at genetic, cellular, organism, and societal scales \shortcite{luengo2011image}.} the molecular scale, the clinical scale, and the societal scale (see Figure \ref{fig:scales}). From a molecular perspective, AI can be used to estimate the structure of SARS-CoV-2-related proteins, identify existing drugs that may be repurposed to treat the virus, propose new compounds that may be promising for drug development, identify potential vaccine targets, improve diagnosis, and better understand virus infectivity and severity. From a clinical perspective, AI can support COVID-19 diagnosis from medical imaging, provide alternative ways to track disease evolution using non-invasive devices, and generate predictions of patient outcomes based on multiple data inputs, including electronic health records.
From a societal perspective, AI has been applied in several areas of epidemiological research that involve modeling empirical data, including forecasting the number cases given different public policy choices. Other works use AI to identify similarities and differences in the evolution of the pandemic across regions. AI can also help investigate the scale and spread of the ``infodemic'' in order to address the propagation of misinformation and disinformation, as well as the emergence of hate speech. 
In addition to this, we review open-source datasets and resources that are available to facilitate the development of AI solutions. Sharing and hosting data and models, whether they be clinical, molecular, or societal, is critical to accelerate the development and operationalization of AI to support the response to the COVID-19 pandemic.

The purpose of this review is not to evaluate the impact of the described techniques, nor to recommend their use, but to show the reader the extent of existing applications and to provide an initial picture and road map of how Artificial Intelligence could help the global response to the COVID-19 pandemic. Based on our review of the literature, we conclude with a series of observations and recommendations. First, we note that while there is a broad range of potential applications of AI covering medical and societal challenges created by the COVID-19 pandemic, few of them are currently mature enough to show operational impact. Second, we recommend that AI solutions targeted at critical application settings -- such as clinical ones -- take into account existing regulatory and quality assurance frameworks to ensure the validity of use and safety, as well as to minimize potential risks and harms. 
Finally, we argue that international AI cooperation based on multidisciplinary research and open science is needed to accelerate the translation of research into global solutions which can be tailored and adapted to local contexts. More detailed discussions of these points can be found in \shortciteA{luengo-oroz2020artificial} and \shortciteA{luccioni2020considerations}.

\section{Article Selection}

\begin{figure}
\centering
 \includegraphics[width=\linewidth]{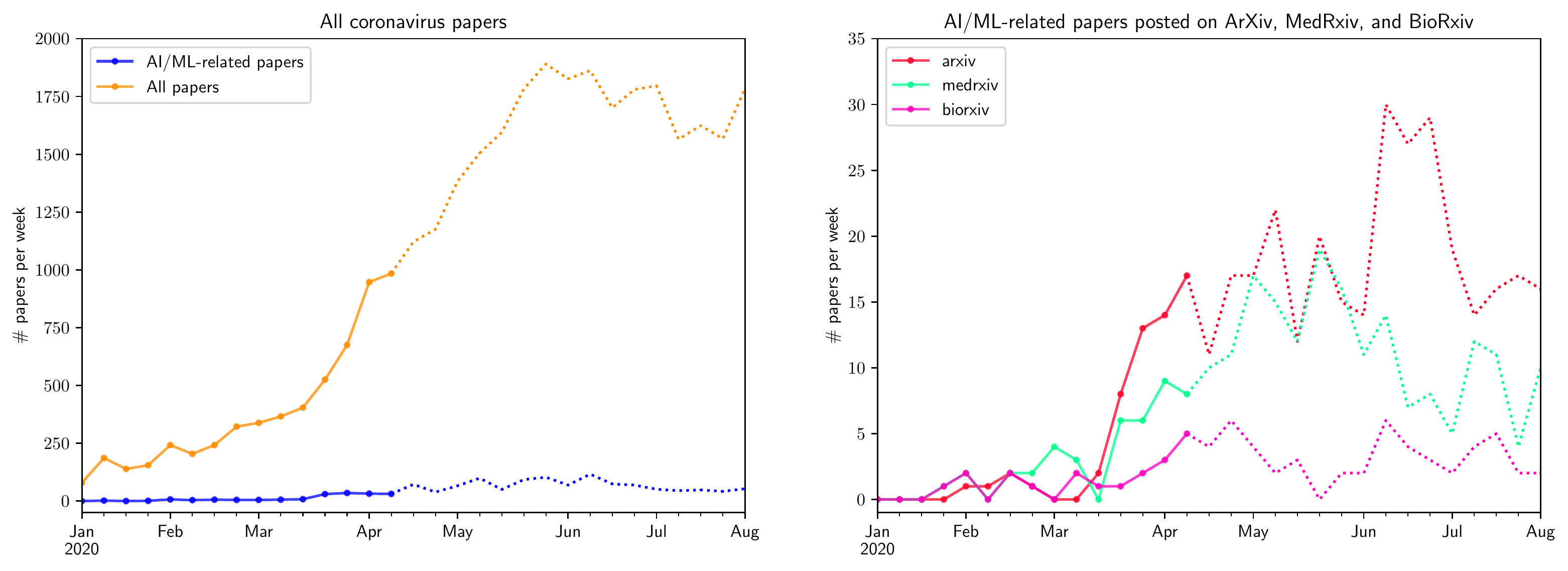}
\caption{Increase over time of scientific articles and preprints related to COVID-19 between January 2 and August 1, 2020. Note that the $y$ axis scales differ between plots.}\label{fig:n_articles}
\end{figure}

The mobilization of the scientific community to address the pandemic is unprecedented in its scale. An automated search for papers posted in the
\urllink{https://pages.semanticscholar.org/coronavirus-research}{COVID-19 Open Research Dataset} (CORD-19) \shortcite{wang_cord-19_2020} identified over 30,000 coronavirus-related papers posted between January 1 and August 1, 2020. Of these papers, over 1,000 included the phrases ``machine learning'', ``artificial intelligence'', ``deep learning'', or ``neural network'' in the title or abstract. As shown in Figure \ref{fig:n_articles}, the number of papers has grown dramatically since mid-March 2020. 

\begin{figure}[h!]
\centering
\includegraphics[width=0.9\linewidth]{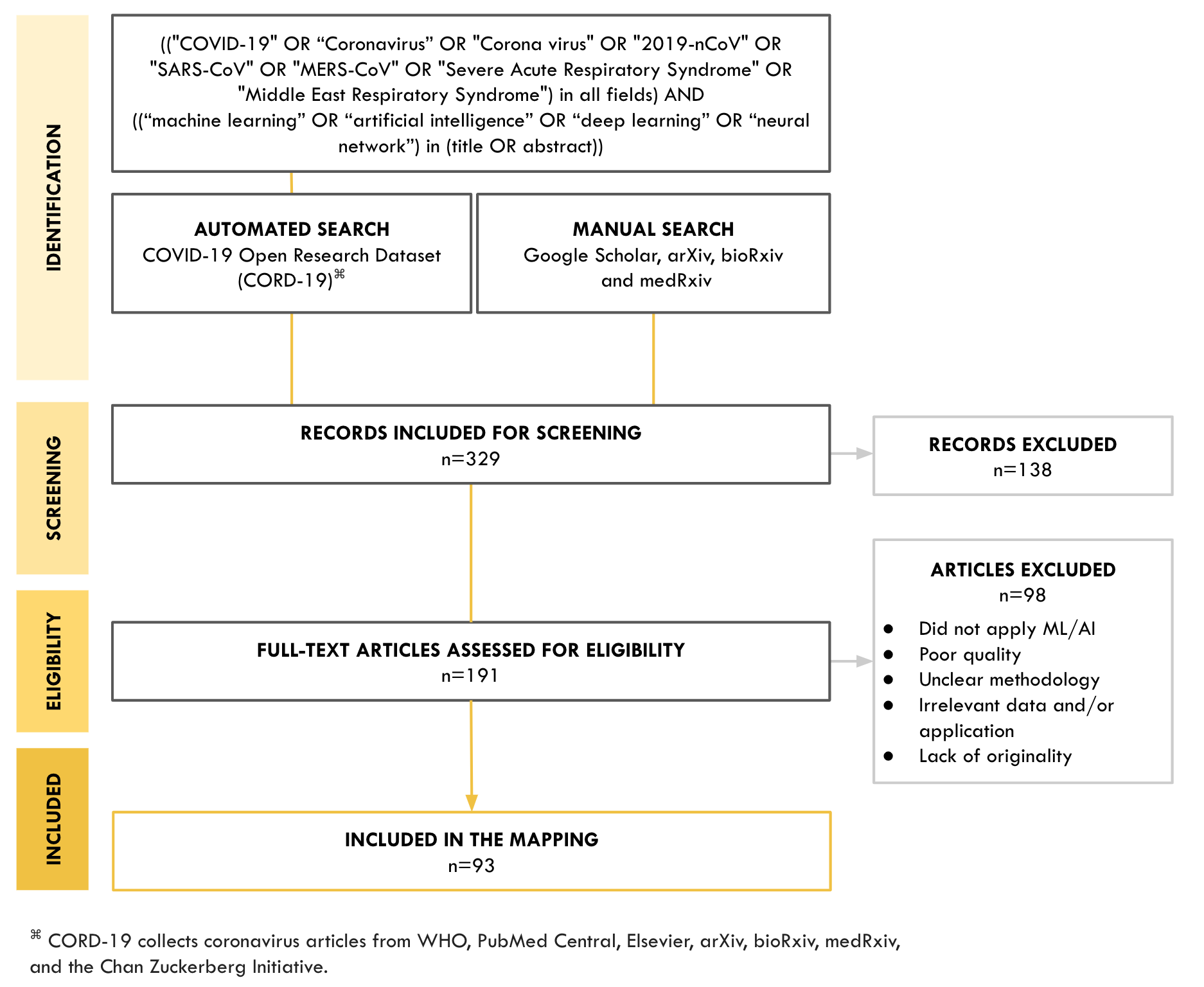}
\caption{Flow chart illustrating how articles were selected for the review, adapted from the PRISMA standard.}\label{fig:prisma}
\end{figure}

To select articles for review, we conducted manual searches of pre-print servers and Google Scholar; we supplemented this with an automated search of the CORD-19 dataset. We focused on manuscripts released between the dates of January 1 and April 10, 2020,\footnote{Given the rapid pace of publication, we note that this survey will not be comprehensive when published. Instead, we have concentrated our efforts on providing a thorough review of publications during the initial response to the pandemic.} and screened articles based on quality, originality, and clarity. The selection process is described further in Figure \ref{fig:prisma}.

The scope of this review is restricted to applications of Machine Learning (ML) and Artificial Intelligence, and we have therefore made judgment calls regarding whether certain methodologies fall into this category. For example, we have included applications where authors have explicitly described the use of models such as neural networks and decision trees, while excluding applications based on simple linear regression models. Furthermore, we note that many of the articles cited are still preprints at the time of writing this review. Some manuscripts have been published since our initial reading of the preprint; where the content of the published article is faithful to that of the initial preprint, we have cited the published article directly.
Given the fast-moving nature of the crisis we strove to be comprehensive in our coverage, but the full scientific rigor of the remaining unpublished manuscripts should still be assessed by the academic community through peer-reviewed evaluation and other quality control mechanisms. For specificity, we signify all preprints with~\preprint. 

Finally, since this article assumes background knowledge of both Machine Learning and the nature of the SARS-CoV-2 virus, we invite our readers to consult~\citeA{raghu2020survey}\preprint~for further explanation regarding the potential of ML for scientific research, and~\citeA{cdcfaq}\preprint~for additional information about the virology, clinical features, and epidemiology of COVID-19. Accessible overviews of SARS-CoV-2 proteins, the infection process, and molecular modeling can be found in \citeA{corum_bad_2020,corum_how_2020} and \citeA{dror_cs279_2019}.


\section{Molecular Scale: From Proteins to Drug Development}
At the most granular scale of the scientific response to COVID-19, biochemistry applications of AI have been used to better understand the proteins involved in SARS-CoV-2 infection and to inform the search for potential treatments. With respect to the virus itself, four types of structural proteins are of interest: nucleocapsid proteins (N), envelope proteins (E), membrane proteins (M), and spike proteins (S) \shortcite{liu_research_2020,zhavoronkov2020potential}\preprint. Also of interest are a number of non-structural proteins (NSPs), which are crucial for viral pathogenesis,
including the 3-chymotrypsin-like (3C-like) protease (also known as 3CLPro, the main protease/MPro, or nsp5) and the papain-like protease (PLpro, part of nsp3). With respect to the virus' human hosts, research has focused on the angiotensin-converting enzyme 2 (ACE2) protein, a receptor that facilitates the virus' entry into host cells~\shortcite{hoffmann2020sars}. Potential applications of AI on this scale include predicting the structure of these associated proteins, identifying existing drugs which may be effective in targeting these proteins, and proposing new chemical compounds for further testing as potential treatments~\shortcite{zhavoronkov_artificial_2018}.

\subsection{Protein Structure Prediction}
Proteins have a 3D structure, which is determined by their genetically encoded amino acid sequence, and this structure influences the role and function of the protein~\shortcite[Chapter~3]{berg_biochemistry_2002}.
Protein structure is traditionally determined through experimental approaches such as X-ray crystallography, but these can be costly and time-consuming. More recently, computational models have been used to predict protein structure~\shortcite{senior_alphafold_2020}\preprint. There are two primary approaches to the prediction task: \textit{template modeling}, which predicts structure using similar proteins as a template sequence, and \textit{template-free modeling}, which predicts structure for proteins that have no known related structure~\shortcite{jumper_computational_2020}\preprint. 

\shortciteA{senior_improved_2020} have developed a system called AlphaFold which focuses on the latter challenge. The AlphaFold model is based on a dilated ResNet architecture~\shortcite{he2016deep,yu_multi-scale_2016} and uses amino acid sequences, as well as features extracted from similar amino acid sequences using multiple sequence alignment (MSA), to predict the distribution of distances and angles between amino acid residues. These predictions are used to construct a ``potential of mean force'' which is used to characterize the protein's shape \shortcite{senior_protein_2019}. This system has been applied to \urllink{https://deepmind.com/research/open-source/computational-predictions-of-protein-structures-associated-with-COVID-19}{predict the structures} of six proteins related to SARS-CoV-2 (the membrane protein, protein 3a, nsp2, nsp4, nsp6, and papain-like protease)~\shortcite{jumper_computational_2020}\preprint.

\shortciteA{heo_modeling_2020}\preprint~also use a dilated ResNet architecture, implemented as part of the transform-restrained Rosetta (\urllink{https://github.com/gjoni/trRosetta}{trRosetta}) pipeline~\shortcite{yang_improved_2020}, to predict the structure of the above proteins as well as proteins ORF6, ORF7b, ORF8, and ORF10.
The trRosetta network has multiple output heads: one which predicts the distances between residues in a protein's amino acid sequence, and others which predict the orientations between these residues as characterized by five different angles. This approach may allow for better performance by jointly learning features that are relevant to predicting both distance and orientation~\shortcite{yang_improved_2020}. \shortciteA{heo_modeling_2020} refine trRosetta and AlphaFold's predicted structures using molecular dynamics simulations and compare the results with structure predictions from a third approach -- \urllink{https://zhanglab.ccmb.med.umich.edu/COVID-19/}{C-I-TASSER}  ~\shortcite{zheng_deep-learning_2019} -- which incorporates nine different methods for contact prediction. While the authors find that the predicted structures generally have a lot of variability between them, there is some consensus for predicted structures of the papain-like protease, part of nsp4, and the M protein. 

\subsection{Drug Repurposing}
In addition to better understanding the structure of key proteins involved in SARS-CoV-2 infection, a number of research efforts have focused on identifying known compounds which might be effective in mitigating infection -- including, potentially, already-approved drugs. We have identified four distinct approaches to this problem, which are facilitated by AI: the construction of biomedical knowledge graphs, the prediction of protein-ligand binding affinities, the modeling of molecular docking, and the analysis of gene expression signatures.

\subsubsection{Biomedical Knowledge Graphs}
Biomedical knowledge graphs are networks capturing the relationships between different entities -- such as proteins and drugs -- to facilitate higher-level exploration of how they connect. \shortciteA{richardson_baricitinib_2020} use this technique to identify Baricitinib, a drug which is commonly used to treat arthritis via inhibition of JAK1/2 kinases, as a promising therapy for COVID-19 because it inhibits the AP2-associated protein kinase 1 (AAK1) enzyme and may, therefore, make it harder for the virus to infect host cells. Related work has described two approaches which potentially inform the graph construction. First, \citeA{segler_planning_2018} describe an approach to mining a structured database of chemical reactions (\urllink{https://www.elsevier.com/solutions/reaxys}{Reaxys}) using a three-part neural network pipeline combined with a Monte Carlo Tree Search approach (3N-MCTS), in order to understand how various compounds are formed hierarchically from reactions between simpler component molecules. Second, \shortciteA{fauqueur_constructing_2019} describe a strategy for mining a scientific article database (\urllink{https://www.ncbi.nlm.nih.gov/pubmed/}{PubMed}) to identify stylized relationships between gene-disease pairs expressed in individual sentences (e.g., ``GENE promotes DISEASE'').

\shortciteA{ge_data-driven_2020}\preprint~describe a similar approach to constructing a knowledge graph connecting human proteins, viral proteins, and drugs using databases that capture the relationships between these entities. The graph is used to predict potentially effective candidate drugs. This list is further refined using a Natural Language Processing (NLP) model, i.e., a Biomedical Entity Relation Extraction (BERE) approach~\shortcite{hong2020novel} applied to the \urllink{https://www.ncbi.nlm.nih.gov/pubmed/}{PubMed} database, filtered for mentions of the candidate drug compounds and associated coronavirus targets. The authors identify a Poly (ADP-Ribose) Polymerase 1 (PARP1) inhibitor, CVL218, as a promising candidate that is already undergoing clinical trials.

\subsubsection{Prediction of Protein-Ligand Binding Affinities}
Other studies attempt to predict protein-ligand binding affinities in order to tackle the drug repurposing problem. Ligands are substances (molecules and ions) which bind with a protein to trigger a signal, such as activation or inhibition. \shortciteA{hu_prediction_2020}\preprint~use a multitask neural network to predict these affinities, selecting a list of 8 SARS-CoV-2 related proteins which they attempt to target using a database of 4,895 drugs. They identify 10 promising drugs, along with their target proteins and binding affinity scores (which indicate the likelihood that the drug will act as an inhibitor). In an attempt to increase model interpretability, they also estimate the precise regions of each target protein where binding is likely to occur. 

In a similar vein, \shortciteA{zhang_deep_2020} use a Dense Fully Convolutional Neural Network (DFCNN) architecture, trained to predict binding behavior on the PDBbind database \shortcite{wang2004pdbbind}, in order to identify potential inhibitors of the 3C-like protease. They develop a homology (template) model of the target protein using its SARS variant, and explore databases of existing compounds (e.g., ChemDiv and TargetMol) as well as tripeptides to find treatments which may be effective at targeting this protein. 
\shortciteA{nguyen_potentially_2020}\preprint~also build a SARS-based homology model of the 3C-like protease, and apply their Mathematical Deep Learning (MathDL) approach 
to identify potential inhibitors for this protease. In particular, their model uses representations of proteins and ligands based on algebraic topology as inputs, and relies on two main datasets: information on 84 SARS coronavirus protease inhibitors from the ChEMBL database \shortcite{
davies2015chembl,mendez2019chembl}, 
and a more general set of 15,843 protein-ligand binding affinities from the PDBbind database \cite{wang2004pdbbind}. They fit two different convolutional neural networks (CNNs)~\shortcite{lecun2015deep} on this dataset -- a pooled (3DALL) CNN which is trained on both datasets together, and a multitask (3DMT) CNN which is trained on each dataset separately \shortcite{nguyen_machine_2020}\preprint. 
Using a consensus between these CNN models, the authors identify a list of 15 promising drug candidates from the DrugBank dataset \shortcite{wishart2018drugbank}.

 Finally, \shortciteA{beck_predicting_2020} use their own Molecule Transformer-Drug Target Interaction (MT-DTI) model  of binding affinities to identify US Food and Drug Administration (FDA) approved antivirals which may be effective in targeting six coronavirus-related proteins. 
 The MT-DTI model ingests string data in the form of simplified molecular-input line-entry system (SMILES) data and amino acid sequences, and applies a text modeling approach that leverages ideas from the BERT algorithm~\shortcite{devlin_bert_2019}. The model identifies drugs that are expected to be effective in targeting each protein studied. \shortciteA{hofmarcher_large-scale_2020}\preprint~likewise apply a text-based approach to SMILES data (ChemAI), which in turn relies on a 
Long Short-Term Memory (LSTM)~\shortcite{lstms1997} model called SmilesLSTM, to screen almost 900 million compounds from the ZINC database \shortcite{sterling2015zinc} for effectiveness in inhibiting the SARS coronavirus 3C-like protease and the papain-like protease. They rank compounds according to predicted inhibitory effects, predicted toxicity, and novelty, and produce a \urllink{https://github.com/ml-jku/sars-cov-inhibitors-chemai}{list of 30,000 candidate compounds}  for screening.

\subsubsection{Molecular Docking}
Another approach to drug repurposing and discovery involves molecular docking. One method for docking is simulation, in which a wide range of candidate ligands interact with a protein in different orientations and conformations, generating a variety of poses (also known as the binding modes -- i.e., the resulting interactions between the ligand and the protein as they bind). The poses are subsequently scored and used to predict the ligand's binding affinity. Since docking approaches such as simulation can be computationally expensive, some research has studied how to make the search more efficient by narrowing the pool of candidates that must be docked. For example, \shortciteA{ton_rapid_2020} develop a Deep Docking (DD) platform that trains a neural network to predict docking scores, which they use to identify a set of 3 million candidate 3C-like protease inhibitors from a set of over 1 billion compounds extracted from the ZINC database \shortcite{sterling2015zinc}. The authors then dock the resulting compounds, presenting \urllink{https://drive.google.com/drive/folders/1xgA8ScPRqIunxEAXFrUEkavS7y3tLIMN}{the top 1,000 results}. On the other hand, \shortciteA{batra_screening_2020}\preprint~train random forests \shortcite{breiman_random_2001}
on SMILES data to predict binding affinity scores which would result from docking simulations and use this approach to select 187 promising molecules to target the coronavirus S-protein and the ACE2 receptor for a final docking simulation. They also identify 19,000 additional candidate compounds in the BindingDB dataset \shortcite{liu2007bindingdb}.

\subsubsection{Gene Expression Signatures}
A fourth approach to drug repurposing involves discovering therapies which have similar effects to other known effective treatments. To this end, \shortciteA{donner_drug_2018} use the LINCS dataset~\shortcite{keenan2018library} of gene expressions from cells targeted by various perturbagens (i.e., chemical or genetic reagents to treat cells and alter intracellular processes). They learn an embedding with a deep neural network classifier that predicts the perturbagen associated with each signature (i.e., the gene expression that is specifically correlated with a biological state of interest, such as therapeutic response). In order to correctly classify signatures associated with the same perturbagen, the learned embedding should abstract away from the noise in the input data and identify core features that are associated with a perturbagen's effect. Their approach can utilize similarity in the learned embedding space to predict pharmacological similarities in structurally different compounds, and hence expand the horizon of drug repurposing. \shortciteA{avchaciov_ai_2020}\preprint~adapt this approach to find drugs that produce gene expression signatures that are similar to the COBP2 gene knockout, which might limit the replication of SARS-CoV-2 based on the gene's role in the replication of the related SARS coronavirus. They list twenty of the most promising drugs, many of which have already been identified as antivirals; since these drugs have been authorized for clinical trials or already approved, the authors argue that their approach could facilitate the rapid discovery of potentially effective therapies.

\subsection{Drug Discovery}
Another line of research attempts to discover entirely new compounds for use in targeting SARS-CoV-2. \shortciteA{zhavoronkov2020potential}\preprint~use a proprietary pipeline to find inhibitors for the 3C-like protease. Their models use three types of input: the crystal structure of the protein, the co-crystalized ligands, and the homology model of the protein. For each input type, the authors fit 28 different models, including Generative Autoencoders~\shortcite{makhzani_adversarial_2016} and Generative Adversarial Networks~\shortcite{goodfellow_generative_2014-1}. The authors explore potential candidates using a reinforcement learning approach with a reward function that incorporates factors such as measures of drug-likeness, novelty, and diversity. Moreover, they confirm that the identified candidate molecules are dissimilar to existing compounds, suggesting that they have indeed found novel candidate drugs.

\shortciteA{tang_ai-aided_2020}\preprint~also apply a reinforcement learning approach to the discovery of compounds that inhibit the 3C-like protease. Specifically, the authors create a list of 284 molecules known to act as inhibitors in the context of SARS. They break these proteins into a series of 316 fragments, which can then be combined using an advanced deep Q-learning network with fragment-based drug design (ADQN-FBDD) that rewards three aspects of discovered molecules: a drug-likeness score, the inclusion of pre-determined ``favorable'' fragments, and the presence of known pharmacophores (which are abstract design patterns believed to be correlated with a compound's effectiveness~\shortcite{qing_pharmacophore_2014}). The 4,922 results are heuristically filtered, and the 47 top compounds are assessed with molecular docking, from which the researchers then select the top most promising compound and manually tailor it to produce suggested variants for testing.

In a third approach, \shortciteA{bung_novo_2020}\preprint~build a generative model to identify potential 3C-like protease inhibitors. Treating SMILES input strings as a time series of characters, they build a classifier that predicts the next character in the string. The model is first trained on 1.6 million molecules from the ChEMBL database \shortcite{
davies2015chembl,mendez2019chembl},
and then adapted to a smaller dataset of protease inhibitors using transfer learning. Reinforcement learning was then used to train the model to generate compounds with desirable properties. After filtering the resulting molecules and docking them, the authors propose 31 candidate inhibitors.

Finally, \shortciteA{nguyen_machine_2020}\preprint~apply a generative network complex (GNC) for drug discovery. Their pipeline involves gated recurrent unit (GRU) based encoders and decoders which ingest SMILES strings and propose new variants with the help of a deep neural network (DNN) between the encoder and decoder that optimizes candidate variants. They also use a pretrained 2D fingerprint-based DNN (2DFP-DNN) as well as their MathDL approach (discussed briefly above) to further predict the properties of the resulting drugs. The authors identify 15 novel candidate drugs and also analyze two proposed HIV drugs to estimate their efficacy against SARS-CoV-2. A similar approach to drug discovery is described in \shortciteA{chenthamarakshan_target-specific_2020}\preprint; the authors' Controlled Generation of Molecules (CogMol) framework uses a variational autoencoder (VAE) trained on SMILES strings to learn molecule embeddings. On these embeddings, the authors train a model to predict drug properties and protein binding affinities and use this to constrain the search for novel strings using Conditional Latent (attribute) Space Sampling (CLaSS). The authors also use a multitask DNN to predict toxicity, in order to avoid proposing candidate drugs with a low probability of success later on in the testing pipeline. The authors focus their search on drugs which target nsp9, the 3C-like protease, and the receptor-binding domain (RBD) associated with the S protein, proposing \urllink{https://covid19-mol.mybluemix.net/}{3,000 of the top candidates} for further study.

In the process of mounting an immune response, B-cells in the body produce antibodies, which attack the part of the pathogen (the virus) known as an antigen. \citeA{magar_potential_2020}\preprint~thus take a different approach to discovering new therapies in which they search for antigen-neutralizing antibodies. They first construct a training dataset (VirusNet) consisting of 1,933 known antigen-antibody sequences from related illnesses such as HIV, SARS, Ebola, and influenza. The authors then train classification models such as XGBoost~\shortcite{chen2016xgboost} on embeddings of the molecular graphs of these antigens and antibodies to predict whether an antibody will have a neutralizing effect on an antigen. Finally, the authors mutate the SARS coronavirus antibody sequence to generate 2,589 candidate antibody sequences. Given the subset of these antibodies which are predicted to be effective by the algorithm, they filter these mutations for valid and stable variants (which they identify through the use of molecular dynamics simulations), and ultimately propose 8 antibodies as potentially effective treatments.

\subsection{Vaccine Discovery}
Another area of interest is vaccine discovery. In addition to producing virus-neutralizing antibodies via B-cells as described above (humoral immunity), the body also uses T-cells to attack the virus directly (cellular immunity). There is a subset of T-cells called memory cells which recognize the antigen of a formerly eliminated pathogen, and can quickly activate more effector T-cells upon re-exposure. These processes inform the targets for vaccine design. As part of the immune response, helper proteins called major histocompatibility complex proteins (MHC I and MHC II proteins) present the binding regions of antigens, called epitopes, for antibodies, B-cells, or T-cells to bind to and attack. These MHC I and MHC II proteins are encoded by Human Leukocyte Antigen (HLA) gene complexes, which vary from person to person. In this context, vaccine design involves two key objectives: (1) identifying suitable epitopes for targeting, and (2) ensuring that these epitopes can be presented by MHC proteins which are produced by different HLA alleles (i.e., variants of a gene) that occur in the population. 

For example, \citeA{fast_potential_2020}\preprint~search for B- and T-cell epitopes. They identify 405 potential T-cell epitopes that can be presented by MHC I and MHC II proteins, as well as two B-cell epitopes on the S-protein. The search for the T-cell epitopes relies on two previously-developed neural networks to predict MHC presentation, NetMHCPan4~\shortcite{jurtz_netmhcpan-4.0_2017} and MARIA~\shortcite{chen_predicting_2019}. Upon identifying potential epitopes, the authors examine 68 different genetic variants of SARS-CoV-2 to study how the virus mutates, and identify parts of the virus that are more or less prone to evolution. They conclude that S-protein epitopes may be a good target for vaccines because they contained no nearby mutations in the sample. In an alternative approach, \shortciteA{ong_covid-19_2020} use their Vaxign-ML framework, which leverages supervised classification models such as XGBoost~\shortcite{chen2016xgboost}, in an effort to predict which viral proteins may serve as effective vaccine targets. While the authors find that the S-protein is the best candidate, they also identify five possible NSPs -- most promisingly, nsp3 and nsp8 -- as good candidates for the vaccine target. 

To the best of our knowledge, three candidate vaccines that reported the use of ML in their development have been approved for clinical evaluation \cite{who2020vaccines}. Nevertheless, it is worth noting that they were created by corporations which published very limited information regarding their methodologies for incorporating ML into their vaccine development pipelines.

\subsection{Improving Viral Nucleic Acid Testing}
Researchers are also applying Machine Learning in an attempt to improve the current viral nucleic acid detection test. \shortciteA{metsky2020CRISPR}\preprint~combine ML with CRISPR (a tool which uses an enzyme to edit genomes by cleaving specific strands of genetic code) to develop 
assay designs for detecting 67 respiratory viruses, including SARS-CoV-2.
The authors note that this technology can speed up the processing of test samples in order to assist overburdened diagnostic facilities, as well as help address the challenge of false positives that occur as a result of sequence similarity between SARS-CoV-2 and other coronaviruses. ML models have been built to rapidly design assays which are predicted to be sensitive and specific, and cover a diverse range of genomes. The authors state that they are aiming to build a Cas13-based point-of-care assay for SARS-CoV-2 in the future. 

\shortciteA{lopez-rincon_accurate_2020}\preprint~take another approach, in which they apply a CNN model \cite{lecun2015deep} to nucleic acid sequences to classify whether they are associated with SARS-CoV-2 and therefore potentially improve the accuracy of diagnosis. They contrast SARS-CoV-2 with other human coronaviruses from the 2019nCoVR repository~\shortcite{zhao_2020_novel}, as well as other genome sequences with the ORF1ab protein from GenBank~\shortcite{benson2012genbank}. The authors use a 21-base pair convolution window over the whole genome and visualize the 
network's convolution and max-pooling layers to understand which particular sequences help to identify SARS-CoV-2. Using only the 21-base-pair sequences retained after the max-pooling layer, they subsequently fit a simpler classification model (e.g., logistic regression) for the original classification task, adding a further layer of feature selection to identify the most predictive sequences. They also apply this classification approach to distinguish between hospitalized and asymptomatic cases. Based on these results, they suggest that a limited number of 21-base-pair sequences might suffice to identify SARS-CoV-2 and predict case severity. Although the data of the study was limited, this work highlights that bioinformatic processes offer opportunities for researchers to improve existing diagnostic tools.

\subsection{Better Understanding Severity and Infectivity}
Additional efforts have used Machine Learning to better understand SARS-CoV-2 infection severity and infectivity (how likely it is that a pathogen can infect a host) using protein sequences. For example, \shortciteA{gussow_genomic_2020} use Support Vector Machines (SVM)~\shortcite{cortes1995support} on genomes from different coronaviruses
to identify which parts of coronavirus' protein sequences distinguish high case fatality rate (high-CFR) from low-CFR variants. \shortciteA{bartoszewicz_interpretable_2020}\preprint~use reverse-complement neural networks built from CNN and LSTM architectures to detect whether a virus has the potential to infect a human host using its viral genome sequence, and apply machine learning interpretability techniques to identify the parts of the sequence that are most associated with infectivity. 
Finally, \shortciteA{randhawa_machine_2020} adopt a Machine Learning with Digital Signal Processing (ML-DSP) approach, which uses supervised learning approaches such as SVM~\shortcite{cortes1995support} and K-nearest neighbors (KNN) to predict the taxonomic classifications of viruses based on their genomic sequences at different levels of the taxonomic hierarchy. Their findings support the classification of SARS-CoV-2 as a sarbecovirus of the betacoronavirus class and the hypothesis that it came from bats. 

\section{Clinical Scale: From Diagnosis to Outcome Predictions} \label{diagnosis}

 To date, most clinical applications of AI for the COVID-19 response have focused on diagnosis based on medical imaging, with an increasing number of studies exploring non-invasive monitoring techniques. In recent literature, we have found several works that use AI to support diagnosis from computational tomography (CT) and X-ray scans, in addition to others that use patient medical data to predict the evolution of the disease, and original non-invasive measurements for monitoring purposes.

\subsection{Medical Imaging for Diagnosis}\label{sec:diagnosis}

Reverse Transcription Polymerase Chain Reaction (RT-PCR) tests are the key approach used for diagnosing COVID-19, however they present limitations in terms of resources, specimen collection, time required for the analysis, and performance~\shortcite{ai2020correlation}. As such, there is growing interest in other diagnostic methodologies that use medical imaging for the screening and diagnosis of COVID-19 cases~\shortcite{kanne2020essentials}. This is notably due to the fact that COVID-19 has been found to exhibit particular radiological signatures and image patterns which can be observed in medical imagery~\shortcite{fang2020sensitivity}, but the identification of these patterns remains time-consuming even for expert radiologists. This makes image analysis from lung CT and X-ray scans of COVID-19 patients a prime candidate for ML-based approaches which could help accelerate the analysis of these scans, although the extent to which imaging can be used for diagnosis is still under discussion~\shortcite{ng2020imaging,weinstock2020}.

There are several approaches that aim to leverage Machine Learning for diagnosing COVID-19 from CT scans, via binary (i.e., healthy vs. COVID-19 positive)~\shortcite{wang2020deep,chen2020deep,gozes2020coronavirus}\preprint~or multi-class (healthy patients vs. COVID-19 vs. other types of pneumonia)~(\shortciteR{song2020deep}\preprint; \shortciteR{xu2020deep,li2020artificial}) classification tasks using neural networks. These approaches use different architectures such as Inception~\shortcite{szegedy2015going}, UNet++~\shortcite{zhou2018unet} and ResNet~\shortcite{he2016deep}, which can be trained directly either on raw CT scans, or on scans labeled with regions of interest identified by radiologists. Some studies also adopt a hybrid approach, combining off-the-shelf software with bespoke ML approaches in order to achieve higher accuracy. For example, in ~\shortciteA{gozes2020rapid}\preprint, a commercial medical imaging program is used for initial image processing and then combined with a custom ML pipeline. This two-step ML approach consists of a U-Net architecture~\shortcite{ronneberger2015} trained on medical images of lung abnormalities in order to pinpoint lung regions of interest and a Resnet-50 architecture~\shortcite{he2016deep} trained on ImagetNet~\shortcite{deng2009imagenet} and then fine-tuned on COVID-19 cases in order to classify the images as COVID-positive or healthy. The resulting architecture is able to both facilitate initial diagnosis and track patient progress by measuring disease severity, and it is used in a tool that has been deployed at hospitals worldwide to help radiologists accelerate the analysis of new cases. 

X-ray images, and specifically chest radiographs, can also be used for COVID-19 detection. Given the accessibility and potential portability of the imaging equipment needed, X-ray images can be an alternative in settings where access to advanced medical equipment such as CT scanners is limited, or while waiting for the results of RT-PCR testing. As shown in \shortciteA{abbas2020classification}\preprint, \shortciteA{bukhari2020diagnostic}\preprint, and \shortciteA{hammoudi2020deep}\preprint, there is potential in the use of Deep Learning approaches on X-ray imagery, using architectures similar to the ones used for CT scans (e.g., ResNet~\shortcite{he2016deep} and CNNs~\shortcite{lecun2015deep}). Some of the existing systems have received certification and have been deployed in hospitals and clinics worldwide~\shortcite{murphy2020covid}. However, further work is ongoing in order to make predictions interpretable~\shortcite{karim2020deepcovidexplainer,ghoshal2020estimating}\preprint~and to ensure that the models can be deployed in mobile and low-resource settings~\shortcite{li2020covidmobilexpert}\preprint.

 Studies which report operational deployment, such as~\shortciteA{shan2020lung}\preprint, have opted for human-in-the-loop diagnostic approaches to reduce the analysis time required with the help of ML architectures. The authors use small manually-labeled batches of data for training an initial model based on the V-Net architecture~\shortcite{milletari2016}. This model then proposes segmentation of new CT scans, which can be corrected by radiologists and fed back into the model in an iterative process.
 This approach has enabled the development of a Deep Learning-based system for both automatic segmentation and the counting of infection regions, as well as assessing the severity of COVID-19, i.e., the percentage of infection in the whole lung. The authors show not only that the model
 improved its own performance incrementally, but also that the human time required for analysis of new images dropped from over 30 minutes initially to under 5 minutes after 200 annotated examples were used to train the model, reducing the effort required by radiologists to review a new scan. This is a promising line of research which harnesses the power of ML alongside human annotation and expertise in a complementary and mutually beneficial manner.
 
 While encouraging results have been achieved by many medical imagery-based AI diagnostic methods, in order for these methods to be used as clinical decision support systems, they should undergo clinical evaluation and comply with regulatory and quality control requirements. In particular, their performance should be validated on a relevant and diverse set of training, validation, and test datasets, and they should demonstrate effectiveness in the clinical workflow~\shortcite{nagendran2020artificial} and adhere to adequate diagnosis reporting guidelines~\shortcite{wynants2020prediction}. We note that most of the papers we reviewed lacked provisions for these measures, relying on small and poorly-balanced datasets with flawed evaluation procedures and no plan for inclusion in clinical workflows. Nonetheless, some commercial analysis tools have been certified for use as medical devices and deployed in hospitals worldwide both for patient diagnosis and outcome prediction.

\subsection{Non-invasive Measurements for Disease Tracking}

There are also a number of approaches that do not require specialized medical imaging equipment for diagnosing and tracking COVID-19. For example, one study used a GRU neural network~\shortcite{cho2014learning} trained on footage from Kinect depth cameras to identify patient respiratory patterns~\shortcite{wang2020abnormal}\preprint, based on recent findings suggesting that COVID-19 generates respiratory patterns which are distinct from those of the flu and the common cold, notably because they exhibit tachypnea (rapid respiration)~\shortcite{cascella2020features}. While these abnormal respiratory patterns are not necessarily perfectly correlated with a real-world diagnosis of COVID-19, prediction of tachypnea could be a relevant first-order diagnostic feature that may contribute to large-scale screening of potential patients. Furthermore, new studies aim to understand how wearable device data can help COVID-19 tracking, based on previous clinical research that has demonstrated the value of aggregated signals from resting heart rates acquired from smart watches for influenza surveillance~\shortcite{radin2020harnessing}.

Finally, a growing number of efforts aim to utilize mobile phones for COVID-19 detection, for instance by using embedded sensors to identify COVID-19 symptoms such as coughing, fatigue and nausea~\shortcite{maghdid2020novel}\preprint, or via phone-based surveys to filter high-risk patients based on responses to key questions regarding travel and symptoms~\shortcite{rao2020identification}. There is also ongoing research on the analysis of recorded cough sounds for preliminary COVID-19 diagnosis~\shortcite{imran2020ai4covid}, which can assist with telemedicine approaches as well as initial triage efforts. While these are important efforts given the ubiquity and accessibility of mobile phone technology, these studies are not sufficiently advanced to evaluate their performance, so more extensive testing and clinical investigations are needed for deployment.

\subsection{Patient Outcome Prediction}

 It is crucial to know which factors can put patients at risk for hospitalization, developing acute respiratory distress syndrome (ARDS), and death from respiratory failure. In this vein, there have been several recent papers that predict potential patient outcomes and propose triage approaches based on features contained in patients’ medical data and blood tests, in order to help clinicians identify high-risk patients and those at risk of later development of ARDS~(\shortciteR{feng2020novel}\preprint; \shortciteR{yan2020prediction}\preprint; \shortciteR{jiang2020towards}). Using approaches such as the XGBoost algorithm~\shortcite{chen2016xgboost} and Support Vector Machines~\shortcite{cortes1995support}, these approaches aim to identify key measurable features to predict mortality risk, which can later be tested for in hospitals upon patient admission and during the hospital stay. Clinical indicators that were identified using these ML-driven approaches include lactic dehydrogenase (LDH), lymphocytes, and high-sensitivity C-reactive protein (CRP)~\shortcite{yan2020prediction}\preprint; alanine aminotransferase (ALT), myalgias, and hemoglobin~\shortcite{jiang2020towards}; and Interleukin-6, Systolic blood pressure, and Monocyte ratio~\shortcite{feng2020novel}\preprint, although more research is needed to define specific thresholds and ranges of these indicators.

Furthermore, several complementary studies aim to also leverage medical imagery for patient
outcome prediction. These include carrying out severity assessment~\shortcite{tang2020severity}, predicting the need for long-term hospitalization based on CT imaging data~\shortcite{qi2020machine}\preprint, and patient risk stratification based on X-ray images~\shortcite{wang2020covid}\preprint. A hybrid approach has also been proposed for this purpose, utilizing both CT findings as well as clinical features to predict the severity of COVID-19~\shortcite{shi2020deep}. The clinical features that were identified in this study, i.e., LDH and CRP, are similar to those identified in the purely clinical studies mentioned above; this overlap is promising for eventual clinical monitoring of these indicators. While these studies are limited both in scope and in data, they constitute important avenues of research that can be complemented and extended with clinical data from incoming cases around the world, thereby hopefully improving the prognosis of all patients and reducing the mortality of those that are critically ill. 

\subsection{Hospital Capacity Planning}

Forecasting hospital occupancy is necessary for preparation, planning, and optimization in overstretched health systems during the COVID-19 pandemic. The availability of Intensive Care Units (ICUs) equipped with artificial respiratory support has proven critical when managing the pandemic, and the lack of equipped ICU beds is one of the main factors that could make the healthcare system collapse in the face of the virus. At this stage, there are a number of research projects which attempt to effectively match existing resources (e.g., ICUs, ventilators, and personal protective equipment) with the incoming demand for care, which fluctuates from one day to the next and can increase drastically in a short period of time. These approaches have used stochastic process simulations to better predict ICU capacity based on the number of fully utilized ICU beds for COVID-19 and non-COVID-19 patients, as well as the rate of incoming COVID-19 patients~\shortcite{alban2020icu}. An AI- based hospital resource optimization tool is expected to be deployed by the National Health Service in the United Kingdom to monitor and predict the upcoming demand for intensive care beds and ventilators needed to treat patients with COVID-19. Other approaches go further, using patient characteristics such as age, gender and co-morbidities as features in a Bayesian optimization algorithm that is able to predict future resource usage at both an individual and a hospital level~\shortcite{alaa2018autoprognosis}.  There are also a number of proprietary and commercial tools that are already used in hospitals and clinics, and are being adapted to this new situation; further validation and methodological details will be required to assess their utility in the context of COVID-19~\shortcite{singh2020validating}\preprint.

\section{Societal Scale: Epidemiology and Infodemiology}

At the highest level of granularity, we review the use of AI at the societal scale. In this section we focus on applications to the fields of epidemiology and infodemiology, finding several parallel approaches in both areas. With respect to epidemiology, many works focus on either supplementing or augmenting classical epidemiological techniques, largely from a predictive standpoint. With respect to infodemics, there has been a similar focus on understanding spread and interaction in the context of information, as well as discussions on proactive actions which can be taken to slow or halt the spread of misinformation.

\subsection{Epidemiology}

The spread of the SARS-CoV-2 virus across the globe has received much policy attention, with advice at the national and local level changing daily in many locations as new information and model forecasts become available. Understanding how the virus is transmitted, and its likely effect on different demographics and geographic locations, is crucial for public health interventions.

The field of epidemiological research is vast, and given the relevance and scale of the pandemic, as well as the new data becoming available, multiple modeling efforts have emerged. While most of these endeavors build on well-established classical models (such as susceptible-infected-recovered (SIR) models) fine-tuned to the COVID-19 situation, we focus here on cases specifically employing Machine Learning techniques for epidemiological modeling tasks.

\subsubsection{Modeling and Forecasting Statistics}

Most AI applications developed for epidemiological modeling have focused on forecasting national and local statistics such as: the total number of confirmed cases, mortality, and recovery rates.
Many authors have attempted to identify optimal approaches or model architectures for understanding and forecasting data. These works employ modeling techniques such as an LSTM-GRU architecture \shortcite{lstms1997,cho2014learning} for time series analysis and prediction \shortcite{dutta2020machine}\preprint, or CNN~\shortcite{lecun2015deep} based approaches in which numerical data has been combined and reshaped into ``images" \shortcite{huang2020multiple}\preprint. In addition, new forecasting models for predicting the total number of confirmed cases have been developed. For example, \shortciteA{alqaness2020optimization} combine an adaptive neuro-fuzzy inference system (ANFIS) \shortcite{jang1993anfis} with an enhanced flower pollination algorithm (FPA) \shortcite{yang2012flower} and salp swarm algorithm (SSA) \shortcite{mirjalili2017salpsa} to optimize the parameters of the model. The robustness of their approach is then assessed by training and testing on weekly confirmed influenza cases collected by the US Centers for Disease Control and the World Health Organization (WHO) over two different four-year periods. 

While these studies show how a range of different architectural choices can be made when building forecasting models, they demonstrate the complexities involved in choosing between such models and the non-trivial interplay between architectures, hyperparameters, and datasets. Moreover, since much of the data collected for COVID-19  modeling tasks is limited, the choice of models and datasets can have significant effects on overall performance. In an attempt to address this, a simple framework has been suggested for exploring models and datasets during testing by ensuring that models of different categories, as defined by the authors, are tested in parallel \shortcite{fong2020finding}. Using this framework, the authors propose and compare a polynomial neural network with corrective feedback (PNN+cf) \shortcite{ivakhnenko1970heuristic} against other model architectures. This model was found to achieve optimal performance in predicting daily statistics on small datasets taken from Chinese health authorities.

Social media and other online data sources also provide a rich source of information for understanding public opinion, perception, and behavior. Such information can be incorporated into modeling efforts to augment existing data with the aim of providing more contextual understanding. For example, \shortciteA{liu2020machine}\preprint~combine related internet search and news media activity with data from the Chinese Center for Disease Control and daily forecasts from GLEAM \shortcite{balcun2010modeling}, an agent-based mechanistic model, in order to produce 2-day forecasts for a range of statistics. The authors first cluster provinces based on geo-spatial similarities in COVID-19 activity, and then train a separate model on each cluster. An existing autoregressive model \shortcite{yang2015argo,lu2019improved} is adapted for forecasting.

In a similar manner, data pertaining to Google search queries and news media have been used as inputs to forecasting models for predicting daily COVID-19-related statistics. \shortciteA{lampos2020tracking}\preprint~assess the frequency of searches for different symptoms against data derived from a UK National Health Service survey of COVID-19 patients in which symptoms were recorded. Using this data, along with prior daily statistics, the authors train an ElasticNet \shortcite{zoue2005elasticnet} model for forecasting future trends. Finally, the authors investigate the transferability of their models between countries. This type of approach could be useful for probing the viability of training a model on data-rich countries and applying it to a data-poor ones, although the results of such a transferred model will have to be tailored to local contexts given possible differences in demographic characteristics and cultural norms.

\subsubsection{Clustering}

Countries have experienced different outbreak timings and growth rates based on a range of factors including: international travel, demographics, socioeconomic factors, health care system characteristics, and policy interventions. By assessing commonalities in virus propagation trends, as well as other country and regional data, it may be possible to cluster countries and regions in order to use data from some areas to predict the outbreak in others. While useful at a high level, a significant limitation of this approach is heterogeneous data collection and reporting in different countries due to testing rates, case tracking efforts, and reporting quality and standards, among others.

\shortciteA{larco2020using} take a simple approach to clustering countries using an unsupervised k-means algorithm. The authors cluster 155 countries using data relating to disease prevalence, average health status, air quality, gross domestic product (GDP), and universal health coverage. They find that their model is able to stratify countries according to the number of confirmed cases, although it cannot do so in terms of the number of deaths or the case fatality rate.

More sophisticated approaches have used the latent features of autoencoders, originally trained to predict infection rates, to identify similar groups of regions or countries. For example, \shortciteA{hu2020artificial}\preprint~have compiled a dataset of accumulated and new confirmed cases in 31 provinces and cities of China. After training a modified autoencoder (MAE) for real-time forecasting of new cases, the authors extract information from the autoencoder's latent variable layers to determine the model's most important features for each analyzed region. These features are then fed into a k-means clustering algorithm which groups similar regions for further analysis. This final step is designed to enable more efficient investigation of regions showing common characteristics of interest. Similarly, \shortciteA{hartono2020generating} has proposed training a Topological Autoencoder (TA), a simplified version of a Soft-supervised Topological Autoencoder \shortcite{hartono2019mixing}, on the number of COVID-19 patients across 240 countries using data collected by the Center for Systems Science and Engineering (CSSE) at Johns Hopkins University. The author then studies the latent variables of the TA to create a 2-dimensional clustering of countries.

\subsubsection{Efficacy of Public Policy}

In attempting to manage the pandemic, many national and local governments have introduced public policy interventions, such as social distancing and the quarantining of individuals showing symptoms of COVID-19. The impacts of these measures may be modeled using agent-based approaches, or by introducing regularizers in differential equations governing statistical interaction models, such as SIR approaches. For instance, \shortciteA{hu2020forecasting} use data from WHO reports to train a modified autoencoder (MAE) to predict the number of cases and deaths on a daily basis. The authors encoded different intervention mechanisms according to their perceived strength, and used this variable as an input to the model. 

\shortciteA{dandekar2020neural}\preprint~adopt a different approach which uses data from Wuhan, China to build on the classical SIR model by adding a time-dependent regularizer to model the number of infected people who are in quarantine. Instead of specifying the form of this function and fitting parameters, the authors use a neural network to learn the ``quarantine strength'', $Q(t)$, based on daily reported statistics, which in turn could help to determine the number of people who are able to infect others as a function of the quarantine strength. While this work is heavily dependent on the available data, and does not differentiate between symptomatic and asymptomatic individuals, the use of neural networks to augment well-understood techniques could serve as a powerful modeling tool.

\subsubsection{Risk Assessment}

The models discussed in the previous sections mainly focused on predicting daily aggregate statistics for different regions or countries. Other work has specifically attempted to forecast the risk of outbreaks, often by reducing aggregate statistical trends into a single risk score, which facilitates interpretation, distills information for rapid analysis, and acts as a precursor to further investigation. However, it is important to note that such a distillation may not be robust to important changes in the underlying data or its coverage, and so should be interpreted with caution by policy makers.

\shortciteA{pal2020neural}\preprint~train an LSTM \shortcite{lstms1997} on variables derived from daily statistics and weather data to predict the long-term duration of the pandemic. In assessing which variables should be included in the model, the authors use an ordinary least-squares regression model to assess the p-value of all candidate features. The output of the LSTM is then used alongside explicit fuzzy rules (based on rates of death, confirmed cases, and recovery) to determine a risk category for the country or region. 

In a similar study, \shortciteA{ronsivalle2020prototype}\preprint~looked at the Inherent Risk of Contagion (IRC), which is defined and calculated by the authors for similar geographic regions based on the acceleration of disease spread. The authors use k-means clustering to identify similar regions based on a non-linear combination of demographic and social characteristics and trained a Fully Connected Network (FCN) on data from Lombardy, Italy to forecast the IRC of the remaining provinces and municipalities of the country. 

A more detailed approach was taken by \shortciteA{ye2020alpha}, who develop a hierarchical community-level risk assessment. Given a location, the proposed $\alpha-$Satellite framework provides risk indices associated with different geographic levels (e.g., state, county, and city). To test this framework, the authors use data from the WHO, the United States Centers for Disease Control, county governments, and other media. They incorporate data on new cases, death rates, confirmed cases, demography, mobility, and social media usage. For regions in which social media data is sparse, the authors use a cGAN \shortcite{mirza2014conditional}\preprint~trained on similar areas to generate synthetic social media content. The authors then attempt to estimate how information at each of the different regional levels impacts the others, as well as how different attributes at each level influence the overall spread of the disease. After building a graph defining relationships between different geographic levels, the authors extract the latent variables from an autoencoder which is designed to aggregate information propagated between different nodes on the graph. The autoencoder plays the role of a dimensionality reduction algorithm to better understand the interplay between different geographic areas and their attributes.

\subsubsection{Bayesian Analysis}

Although Bayesian analysis techniques are sometimes considered to be statistical rather than Machine Learning approaches, they can provide useful insights with respect to uncertainty and the handling of small datasets. 
In one study, \shortciteA{roy2020bayesian}\preprint~develop a time-varying Bayesian autoregressive model for counts (TVBARC) with a linear link function to estimate time-dependent coefficients which could allow for better temporal modeling of the virus spread.

A more case-specific application of such methods is employed by \shortciteA{mizumoto2020asymptomatic}, who seek to understand the rate of asymptomatic cases using data on 634 confirmed cases collected during the COVID-19 outbreak on board the \textit{Diamond Princess} cruise ship. The authors use a Bayesian time-series model, with a 
Hamiltonian Monte Carlo (HMC) algorithm and a No-U-Turn-Sampler \shortcite{homan2014nouturn} for model parameter estimation, in order to estimate the probability that a given patient is asymptomatic conditional on infection, along with the duration for which an individual is infected. The authors conclude that 17.9\% of patients are asymptomatic. Although it is unclear if this result applies to the broader population,
contained environments such as this one can be useful for tracking infection because they can allow for more comprehensive case data collection.

\subsection{Infodemiology}

 The WHO defines an infodemic as ``an over-abundance of information – some accurate and some not – that makes it hard for people to find trustworthy sources and reliable guidance when they need it", and deems it a second ``disease" which needs fighting \shortcite{who2020infordmic}. In this section, we highlight efforts to quantify the spread of information surrounding the pandemic and to understand its dynamics. Handling this vast amount of information requires the development and adoption of new tools, particularly for studying the dissemination of misinformation and disinformation. While much AI and ML research has already been carried out in this area, there is still a need for greater understanding of the underlying social dynamics specific to the pandemic. 

Social media and online platforms have become key distribution channels for information surrounding the virus. Although national and international organizations have used these platforms to constructively communicate with the public, populations can also become overwhelmed with information, and the propagation of misinformation and disinformation is increasingly prevalent.

Furthermore, we note that the infodemic may even extend to scientific research. As highlighted in 
Figure \ref{fig:n_articles}, there has been a significant increase in the number of scientific articles related to the SARS-CoV-2 virus. Given that the virus is still relatively new and our understanding is quickly developing, many of these articles are disseminated via preprint archives, making it difficult to assess their quality. This does not mean that information contained within these articles cannot be valuable, but rather that there is a need for ongoing efforts to distill and critically assess this vast body of literature.

\subsubsection{Spread and Interaction}

Understanding more about the dissemination of information is crucial to intervening proactively or reactively. While there is a wealth of literature on information propagation, network analysis, and social media interaction, in this section we specifically discuss those works applying such methods to the current infodemic.

At a high level, some research looks at global trends on Twitter by country. \shortciteA{singh2020first}\preprint~analyze tweet volume according to specific themes discovered in coronavirus-related queries. The authors also analyze posts pertaining to specific myths surrounding the virus, examining the number of tweets containing certain terms they deem related to the myths, as well as the website links included in the tweets (categorized as either high-quality or low-quality sources).

In an effort to find early warning signals of a country or region experiencing an infodemic, \shortciteA{gallotti2020assessing}\preprint~analyze social media posts on Twitter across 64 languages. The authors develop an Infodemic Risk Index (IRI) to quantify the rate at which a given generic user from a country or region is exposed to unreliable posts from different classes of users, i.e., verified humans, unverified humans, verified bots, and unverified bots. The IRI considers the expected number of followers of the focal users which fall into each class, the number of messages the focal users post, and their reliability (as measured by fact-checked samples of the user posts). This study highlights potentially actionable insights, observing that ``the escalation of the epidemics leads people to progressively pay attention to more reliable sources thus potentially limiting the impact of the infodemics," while ``the actual speed of adjustment may make a major difference in determining the social outcome".

In a broad-ranging study, \shortciteA{cinelli2020infodemic}\preprint~analyze interaction and engagement with COVID-19-related social media content. From a collection of eight million comments and posts selected from Twitter, Instagram, YouTube, Reddit, and Gab using COVID-19-related keywords, the authors estimate engagement and interest in COVID-19 and comparatively assess the evolution of discourse on each platform. Interaction and engagement are measured using the cumulative number of posts and the number of reactions to these posts (e.g., comments, likes etc.) across a 45-day period. The authors then employ phenomenological \shortcite{fisman2013idea} and classical SIR models to characterize the reproduction numbers of the posts. Specifically, they examine the average number of secondary cases (users that start posting about COVID-19) created by an ``infectious" (already posting) individual on each of the social media platforms. As in epidemiological models, the authors simulate the likelihood of an infodemic, in which discussion of COVID-19 will grow exponentially in its initial stages. Moreover, the authors examine the spread of misinformation (which they identify using external fact-checking organizations) and find that information from both reliable and less-reliable sources propagates in similar patterns, but that user engagement with posts from the latter is lower across major social media streams.

Similarly, \shortciteA{mejova2020advertisers}\preprint~have examined the use of Facebook advertisements with content related to the virus. The authors used the Facebook Ad Library 
to search for all advertisements using the keywords ``coronavirus" and ``covid-19" and collected results across 34 countries, with most in the US (39\%) and the EU (Italy made up 25\% of the advertising market). While the majority of advertisements were paid for by non-profits to disseminate information and solicit donations, the authors found that around 5\% of advertisements contained possible errors or misinformation. 

\subsubsection{Hate Speech}

Along with the propagation of misinformation and disinformation, the increase in hate speech in recent months has been of significant concern. As reported by the United Nations, there is an alarming rise in verbal abuse which might turn into physical violence against vulnerable and discriminated groups \shortcite{UNhatespeech}.

\shortciteA{velasquez2020hate}\preprint~take a high-level approach to understanding the spread of hateful and malicious COVID-19 information and content within a variety of different social media channels, and attempt to characterize the methods by which such content moves between them. Concerningly, the authors find that hateful content is rapidly evolving and becoming increasingly coherent as time continues. As in \shortciteA{cinelli2020infodemic}\preprint, this study makes a comparison to the epidemiological reproduction number, $R0$, in an attempt to determine the ``tipping point'' at which information will spread more rapidly between information channels.

\shortciteA{schild2020go}\preprint~examine the emergence of Sinophobic behavior on social media, specifically Twitter and 4chan. This study uses data from October 2019 to March 2020 and uses word embeddings to assess context and word similarity over the entire five month period, as well as on a weekly basis. The authors also compared their findings to models trained on content gathered prior to COVID-19. The authors observe a distinct increase in Sinophobic content across social media channels, and conclude that the Web is being ``exploited for disseminating disturbing and harmful information, including conspiracy theories and hate speech targeting Chinese people".

Understanding and fighting the spread of hate speech is of vital importance for the protection of human rights, in particular those of the most vulnerable and marginalized. By better comprehending the dynamics and the landscape of hateful speech, effective intervention mechanisms can be designed to disrupt and change the narrative.

\subsubsection{Positive Action}

In the process of studying the features and dynamics of the infodemic, many of the works mentioned above suggest possible intervention options. In this section we explore several examples of such positive actions that are being considered and/or deployed to counter the infodemic.

The World Health Organization has taken steps to proactively confront the infodemic and bring together actors to assess aspects of the infodemic which still need to be addressed. Indeed, the WHO has been combating the infodemic through the use of its Information Network for Epidemics (EPI-WIN) platform for sharing information with key stakeholders, and is also working with social media and internet search companies to track the spread of specific rumors and ensure that WHO content is displayed at the top of searches for terms related to the virus \shortcite{zarocostas2020lancet}. Indeed, in April 2020, the WHO conducted a wide-ranging consultation on understanding and managing the infodemic \shortcite{who2020infordmic}.

Efforts are also underway to curate specific news content related to the virus and to perform both manual and automated fact-checking and relevance analysis. For instance,  \shortciteA{pandey2020wash}\preprint~have developed a pipeline for assessing the similarity between daily news headlines and WHO recommendations. The pipeline uses word embedding and similarity metrics, such as cosine similarity, to assess the level of relevance between WHO recommendations and news articles. If the similarity is above a certain threshold, the article is displayed on the user's timeline with the associated relevant WHO recommendation. The similarity threshold is determined by human reviewers prior to release and then can be updated through user feedback. 
In the face of conflicting information, such methods could help identify accurate and trustworthy news articles which highlight important guidelines and promote official recommendations.

Another possible intervention strategy under consideration is the use of chatbots, which can be used to disseminate information while relieving pressure on other communication channels, such as question-and-answer hotlines. For example, the WHO who has developed an interactive chatbot in multiple languages that allows users to explore pre-coded topics \shortcite{who2020chatbot}. Finally, there is also the potential to use digital personal assistants to interactively disseminate official information, although governments and international actors would need secure ways to update recommendations as their understanding of COVID-19 changes over time.

\section{Datasets and Resources}

The success of the global effort to use AI techniques to address the COVID-19 pandemic hinges upon sufficient access to data. Machine Learning, and Deep Learning in particular, requires notoriously large amounts of data and computing power in order to develop and train new algorithms and neural network architectures. In this section, we describe some of the datasets and data collection efforts that exist at the present time.

\subsection{Case Data}

The current number and location of cases is essential for tracking the progress of the COVID-19 pandemic, calculating the growth rate of new infections, and observing the impact of preventive measures. Several datasets from organizations such as the WHO~\citeyear{who2020sitrep} and national Centers for Disease Control (CDCs) exist for this purpose. They have been aggregated into public repositories hosted by institutions such as the Johns Hopkins CSSE~\shortcite{dong2020interactive} or on platforms such as GitHub~\shortcite{xu2020epidemiological}, in order provide daily information on COVID-19 cases gathered from a variety of reliable sources. There are also other complementary data sources -- including regional data on school closures, bank interest rates, and even community perceptions of the virus -- which are continuously being added to a data portal hosted by the Humanitarian Data Exchange~\citeyear{humxchange}.
A multitude of algorithms can be applied on this kind of data, including time series forecasting approaches such as LSTM networks~\shortcite{lstms1997} or Autoregressive Integrated Moving Average (ARIMA) models to predict the evolution of cases on a global and regional scale.

There is also an increasing quantity of tools and resources developed specifically for medical professionals and institutions, using data to help them prepare for managing the pandemic. For instance, CHIME~\shortcite{chime} is an open-source COVID-19 Hospital Impact Model for Epidemics based on SIR modeling, which uses the number of susceptible, infectious, and recovered individuals to compute the theoretical number of people infected over time, predict outcomes in specific circumstances, and plan for the quantity of hospital beds that may be needed. While the CHIME project does not currently use ML techniques, it could benefit from them in order to incorporate more features and data points such as hospital capacity information~\shortcite{covidcare}, to be used in applications such as dynamic ventilator allocation and surge capacity planning. Finally, there are also efforts underway to use de-identified, large-scale data to assess mobility changes and their impact on the local evolution of the epidemic, for instance in Italy~\shortcite{pepe2020covid} and in North America~\shortcite{foottraffic}.

\subsection{Text Data}

Unstructured textual data can be an immensely useful source of information, and NLP approaches can be used to mine such data to extract relevant passages and topics. These methods have been applied to a variety of data sources, including scientific articles, news articles, and social media data, in order to 
provide insights and indicators useful to a range of different stakeholders.

\subsubsection{Scientific Literature}
As mentioned in previous sections, Machine Learning approaches can be used to analyze and parse the vast quantity of written information on COVID-19 and other coronaviruses, in order to make it easier for researchers and clinicians to use this information. Key questions of interest include: 
\begin{enumerate}
 \item What is known about the virus' transmission, incubation, and environmental stability?
 \item What do we know about COVID-19 risk factors?
 \item What do we know about non-pharmaceutical interventions?
 \item What do we know about vaccines and therapeutics?
 \item What has been published about ethical and social science considerations?
 \item What has been published about best practices for medical care?
\end{enumerate}

These questions can be studied using different sources, including the WHO Global Research Database on COVID-19~\shortcite{who2020database}\preprint, a curated literature hub for COVID-19 scientific information, and the CORD-19 dataset~\shortcite{wang_cord-19_2020}\preprint, which is currently the largest open dataset available with over 52,000 relevant research articles. Several studies aiming to analyze this information have already been published, including \citeA{ahamed2020information}\preprint, which uses a graph-based model to search through abstracts to find relevant information, and \shortciteA{fister2020discovering}\preprint, which extracts key terms and compares their usage  in pre- and post-COVID-19 articles. There are also several ongoing Kaggle challenges involving this data, with dozens of questions submitted daily and many teams involved. Other scientific research datasets that can be exploited include LitCOVID~\shortcite{chen2020keep}\preprint~and the Dimensions AI Dataset~\shortcite{dimensions}\preprint, which can contain important supplementary information such as clinical trial data when available. Using any of the sources mentioned above, NLP techniques can be applied to develop text mining tools and resources that can help the medical community find answers to key scientific questions regarding the nature and progress of COVID-19. 

\subsubsection{News and Social Media Data}
Depending on the research questions addressed, data from scientific articles can also be complemented with data from other sources, such as news articles and social media. Datasets such as the COVID-19 TweetIDs dataset~\shortcite{chen2020tracking} and the Covid-19 Twitter dataset~\shortcite{banda2020large}\preprint, which are maintained with general coronavirus-related
tweets, can be useful for tracking the propagation of misinformation and unverified rumors on Twitter~\shortcite{chen2020covid19}, as well as for monitoring the reactions of different populations to the virus. A potentially complementary source of information for this task is the COVID-19 Real World Worry Dataset~\shortcite{kleinberg2020measuring}\preprint, which includes labeled texts of individuals' emotional responses to COVID-19, and therefore can contain data regarding public sentiment and the impacts of the pandemic on mental well-being in different regions of the world. 

There is also information available from official sources and news outlets, since the global media coverage of the pandemic is substantial and ongoing. For instance, the Institutional and News Media Tweet Dataset~\shortcite{yu2020open}\preprint~brings together tweets based on a list of manually verified sources, and can be used to track official and institutional messaging around the pandemic in different countries. Repositories such as the Coronavirus News Article database~\shortcite{archive}\preprint~and the COVID-19 Television Coverage Dataset~\shortcite{tv}\preprint~can also be used to explore the question of how both print and television media outlets are covering the outbreak. These are rich sources of data for researchers interested in analyzing how media coverage evolves as the virus spreads globally, or in tracking misleading reports and disinformation in the media.

\subsection{Biomedical Data}
In addition to case data and textual data, biomedical data is an important source of data for training many of the ML algorithms above. 
Biomedical data 
at different scales, from the clinical scale to the molecular scale, can be used to ensure that the ML approaches developed take into account both the different features of patients in different contexts, as well as the evolution of the virus across time and space, the structure of the virus, and the effectiveness of potential therapies.

\subsubsection{Clinical Data}

At this time, there are not many open-source datasets and models that can be used for diagnostic purposes. Some of the CT scan detection approaches described in Section~\ref{sec:diagnosis} are available online and accessible to the public, for instance those of \citeA{wang2020deep} and \citeA{song2020deep}. However, the data used to train the various models described is not systematically shared, although such sharing would be of great value to the ML research community. Several initiatives exist to crowdsource and open source relevant data, for instance the Covid Chest X-ray Dataset~\shortcite{cohen2020covid} for medical imagery and the COVID-19 Risk Calculator~\citeyear{calculator}\preprint~for symptoms, but these are challenging to assemble and maintain manually. Furthermore, while data collection and ML model training can be carried out by computer scientists, data labeling, vetting, and annotation often require the involvement of medical professionals such as radiologists or clinicians. 

To address this lack of accessible data, there is an increasing number of initiatives and repositories that aim to share data and models; for instance, the COVID-19 Dataset Clearinghouse~\citeyear{clearinghouse}\preprint~has links to dozens of open-source data repositories from different geographical areas and levels of granularity. Initiatives such as United Against COVID-19 are particularly important, since they have become online platforms where data scientists and ML researchers can apply their skills to address requests for help from the research community, for instance by performing cleaning of clinical data, extracting actionable information regarding COVID-specific research questions, and collaborating to develop tools for deployment on the ground in hospitals. Such initiatives are promising given their potential to bridge the gap between those with medical and biological knowledge or experience, and those with computational and data skills.

\subsubsection{Molecular Data}
In terms of genomic sequencing and drug discovery, there are several datasets available from pre-existing initiatives, or which have been created from scratch for COVID-19 specifically. On the one hand, tracking the genome sequence of SARS-CoV-2 is crucial for designing and evaluating diagnostic tests, tracing the pandemic, and identifying the most promising intervention options. Notably, the \urllink{https://www.gisaid.org/epiflu-applications/next-hcov-19-app/}{GISAID Initiative}, founded over a decade ago for the specific purpose of promoting the international sharing of influenza virus sequences and related clinical and epidemiological data, is tracking the genomic epidemiology of SARS-CoV-2. Other projects such as Nextstrain~\shortcite{hadfield2018nextstrain} are looking at the genetic diversity of coronaviruses, in order to characterize the geographic spread of COVID-19 by inferring the lineage tree of hundreds of publicly shared genomes of SARS-CoV-2. 

On the other hand, in terms of drug discovery, there are well-established initiatives such as the RCSB Protein Data Bank~\shortcite{burley2017protein} and the Global Health Drug Discovery Institute Portal~\shortcite{GHDDI}\preprint~, which have created centralized portals with data and resources for better understanding COVID-19 and for carrying out structure-guided drug discovery. In addition, CAS, a division of the American Chemical Society, has recently released the open-source COVID-19 antiviral candidate compounds dataset~\shortcite{CAS}\preprint, containing information regarding antiviral compounds and molecules that have similar chemical structures to existing antivirals, to help the discovery of both new and repurposed treatments against the disease. Finally, another potentially interesting crowdsourced resource is the citizen science game \urllink{https://fold.it/}{Fold.it}, which leverages collective intelligence against COVID-19 by challenging participants to design an antiviral protein.

\section{Discussion}

This research mapping exercise suggests that ML and AI can support the response against COVID-19 in a broad set of domains. In particular, we have highlighted emerging applications in drug discovery and development, diagnosis and clinical outcome prediction, epidemiology, and infodemiology. However, we note that very few of the reviewed systems have operational maturity at this stage. In order to operationalize research, it is crucial to define a research road map and a funnel for AI applications to understand how the technology of interest can immediately assist with the response, how it might help later on in the evolution of the current pandemic, and how it can be used to combat future pandemics. In the face of overstretched health care networks, we must strengthen our health systems to sustain services beyond the control and management of COVID-19 in order to truly protect the vulnerable, such as people living with noncommunicable diseases (NCDs). 
As members of a global community of researchers and data scientists, we identify three key calls for action. 

First, we believe that scalable approaches to data and model sharing using open repositories will drastically accelerate the development of new models and unlock data for the public interest. Global repositories with anonymized clinical data, including medical imaging and patient histories, can be of particular interest in order to generate and transfer knowledge between medical institutions. 
To facilitate the sharing of such data, clinical protocols and data sharing architectures will need to be designed and data governance frameworks will need to be put in place. It is important to ensure that research with medical data is subject to strong regulatory requirements and privacy-protecting mechanisms. In particular, clinical AI applications should demonstrate not only their performance on test datasets, but also their effectiveness and safety when integrated into real clinical workflows. Overall, any AI application should undergo an assessment to ensure that it complies with ethical principles and, above all, respects human rights. 

Second, the multidisciplinary nature of the research required to deploy AI systems in this context calls for the creation of extremely diverse, complementary teams and long-term partnerships. Beyond the examples shown in this review, other promising domains in which AI could be used to fight against COVID-19 include robotics (e.g., cleaning or disinfecting robots) and logistics (e.g., the allocation and distribution of personal protective equipment). Funding opportunities 
which encourage such collaborations and define key research directions may help accelerate the success of such partnerships.

Third, we believe that open science and international cooperation can play an important role in this pandemic that knows no borders \shortcite{luengo-oroz2020artificial}. Proven solutions can be shared globally and adapted to other contexts and situations, prioritizing those solutions that target local unmet needs. In particular, given that many international organizations, private sector companies and AI partnerships operate across international borders, they may be in the position to facilitate the knowledge dissemination and capacity building of national health systems. Regions with less capacity can benefit from global cooperation and concentrate their efforts on the most important local challenges. AI systems, methods, and models can act as a compact form of knowledge sharing which can be used in and adapted to other contexts if they are designed to be widely deployable, requiring low energy and computing resources.

We acknowledge the difficulty of adding value through AI in the current situation. Nevertheless, we hope that this review acts as a first step towards helping the AI community understand where it can be of value, which are the promising domains for collaboration, and how research agendas can be best directed towards action against this or the next pandemic.\\

\newpage
\acks{United Nations Global Pulse is supported by the Governments of Sweden and Germany and the William and Flora Hewlett Foundation. JB also is supported by the UK Science and Technology Facilities Council (STFC) grant number ST/P006744/1. AL is supported by grants from IVADO and Mila institutes. Thank you to our colleagues from the M.Tyers Laboratory (IRIC) for their advice.}

\vskip 0.2in
\bibliography{bibliography}

\begin{thebibliography}{}

\bibitem[\protect\BCAY{Abbas, Abdelsamea,\ \BBA\ Gaber}{Abbas
  et~al.}{2020}]{abbas2020classification}
Abbas, A., Abdelsamea, M.~M., \BBA\ Gaber, M.~M. \BBOP2020\BBCP.
\newblock \BBOQ Classification of {COVID}-19 in chest {X}-ray images using
  {DeTraC} deep convolutional neural network\BBCQ\
\newblock {\Bem arXiv preprint arXiv:2003.13815}.

\bibitem[\protect\BCAY{Ahamed\ \BBA\ Samad}{Ahamed\ \BBA\
  Samad}{2020}]{ahamed2020information}
Ahamed, S.\BBACOMMA\  \BBA\ Samad, M.~D. \BBOP2020\BBCP.
\newblock \BBOQ Information mining for {COVID}-19 research from a large volume
  of scientific literature\BBCQ\
\newblock {\Bem arXiv preprint arXiv:2004.02085}.

\bibitem[\protect\BCAY{Ai, Yang, Hou, Zhan, Chen, Lv, Tao, Sun,\ \BBA\ Xia}{Ai
  et~al.}{2020}]{ai2020correlation}
Ai, T., Yang, Z., Hou, H., Zhan, C., Chen, C., Lv, W., Tao, Q., Sun, Z., \BBA\
  Xia, L. \BBOP2020\BBCP.
\newblock \BBOQ Correlation of chest {CT} and {RT-PCR} testing in coronavirus
  disease 2019 ({COVID}-19) in {C}hina: a report of 1014 cases\BBCQ\
\newblock {\Bem Radiology}, 200642.

\bibitem[\protect\BCAY{{Al-qaness}, Ewees, Fan,\ \BBA\ Abd El~Aziz}{{Al-qaness}
  et~al.}{2020}]{alqaness2020optimization}
{Al-qaness}, M. A.~A., Ewees, A.~A., Fan, H., \BBA\ Abd El~Aziz, M. A.~E.
  \BBOP2020\BBCP.
\newblock \BBOQ Optimization method for forecasting confirmed cases of
  {COVID}-19 in {C}hina\BBCQ\
\newblock {\Bem Journal of Clinical Medicine}, {\Bem 9\/}(3), 674.

\bibitem[\protect\BCAY{Alaa\ \BBA\ van~der Schaar}{Alaa\ \BBA\ van~der
  Schaar}{2018}]{alaa2018autoprognosis}
Alaa, A.~M.\BBACOMMA\  \BBA\ van~der Schaar, M. \BBOP2018\BBCP.
\newblock \BBOQ Autoprognosis: Automated clinical prognostic modeling via
  bayesian optimization with structured kernel learning\BBCQ\
\newblock {\Bem arXiv preprint arXiv:1802.07207}.

\bibitem[\protect\BCAY{Alban, Chick, Dongelmans, Vlaar, Sent,\ \BBA\
  Group}{Alban et~al.}{2020}]{alban2020icu}
Alban, A., Chick, S.~E., Dongelmans, D.~A., Vlaar, A.~P., Sent, D., \BBA\
  Group, S. \BBOP2020\BBCP.
\newblock \BBOQ {ICU} capacity management during the {COVID-19} pandemic using
  a process simulation\BBCQ\
\newblock {\Bem Intensive Care Medicine}, {\Bem 46\/}(8), 1.

\bibitem[\protect\BCAY{Avchaciov, Burmistrova,\ \BBA\ Fedichev}{Avchaciov
  et~al.}{2020}]{avchaciov_ai_2020}
Avchaciov, K., Burmistrova, O., \BBA\ Fedichev, P. \BBOP2020\BBCP.
\newblock \BBOQ {{AI}} for the repurposing of approved or investigational drugs
  against {{COVID}}-19\BBCQ\
\newblock {\Bem ResearchGate preprint}.

\bibitem[\protect\BCAY{Balcan, Gonçalves, Hu, Ramasco, Colizza,\ \BBA\
  Vespignani}{Balcan et~al.}{2010}]{balcun2010modeling}
Balcan, D., Gonçalves, B., Hu, H., Ramasco, J., Colizza, V., \BBA\ Vespignani,
  A. \BBOP2010\BBCP.
\newblock \BBOQ Modeling the spatial spread of infectious diseases: the global
  epidemic and mobility computational model\BBCQ\
\newblock {\Bem Journal of Computational Science}, {\Bem 1\/}(3), 132--145.

\bibitem[\protect\BCAY{Banda, Tekumalla, Wang, Yu, Liu, Ding,\ \BBA\
  Chowell}{Banda et~al.}{2020}]{banda2020large}
Banda, J.~M., Tekumalla, R., Wang, G., Yu, J., Liu, T., Ding, Y., \BBA\
  Chowell, G. \BBOP2020\BBCP.
\newblock \BBOQ A large-scale {COVID}-19 {Twitter} chatter dataset for open
  scientific research -- {A}n international collaboration\BBCQ\
\newblock {\Bem arXiv preprint arXiv:2004.03688}.

\bibitem[\protect\BCAY{Bandyopadhyay\ \BBA\ Dutta}{Bandyopadhyay\ \BBA\
  Dutta}{2020}]{dutta2020machine}
Bandyopadhyay, S.~K.\BBACOMMA\  \BBA\ Dutta, S. \BBOP2020\BBCP.
\newblock \BBOQ Machine learning approach for confirmation of {COVID}-19 cases:
  Positive, negative, death and release\BBCQ\
\newblock {\Bem medRxiv preprint medRxiv:2020.03.25.20043505}.

\bibitem[\protect\BCAY{Bartoszewicz, Seidel,\ \BBA\ Renard}{Bartoszewicz
  et~al.}{2020}]{bartoszewicz_interpretable_2020}
Bartoszewicz, J.~M., Seidel, A., \BBA\ Renard, B.~Y. \BBOP2020\BBCP.
\newblock \BBOQ Interpretable detection of novel human viruses from genome
  sequencing data\BBCQ\
\newblock {\Bem bioRxiv preprint bioRxiv:2020.01.29.925354v2}.

\bibitem[\protect\BCAY{Batra, Chan, Kamath, Ramprasad, Cherukara,\ \BBA\
  Sankaranarayanan}{Batra et~al.}{2020}]{batra_screening_2020}
Batra, R., Chan, H., Kamath, G., Ramprasad, R., Cherukara, M.~J., \BBA\
  Sankaranarayanan, S. \BBOP2020\BBCP.
\newblock \BBOQ Screening of {{Therapeutic Agents}} for {{COVID}}-19 using
  {{Machine Learning}} and {{Ensemble Docking Simulations}}\BBCQ\
\newblock {\Bem arXiv preprint arXiv:2004.03766}.

\bibitem[\protect\BCAY{Beck, Shin, Choi, Park,\ \BBA\ Kang}{Beck
  et~al.}{2020}]{beck_predicting_2020}
Beck, B.~R., Shin, B., Choi, Y., Park, S., \BBA\ Kang, K. \BBOP2020\BBCP.
\newblock \BBOQ Predicting commercially available antiviral drugs that may act
  on the novel coronavirus (2019-{{nCoV}}), {{Wuhan}}, {{China}} through a
  drug-target interaction deep learning model\BBCQ\
\newblock {\Bem Computational and Structural Biotechnology Journal}, {\Bem 18},
  784--790.

\bibitem[\protect\BCAY{Benson, Cavanaugh, Clark, Karsch-Mizrachi, Lipman,
  Ostell,\ \BBA\ Sayers}{Benson et~al.}{2012}]{benson2012genbank}
Benson, D.~A., Cavanaugh, M., Clark, K., Karsch-Mizrachi, I., Lipman, D.~J.,
  Ostell, J., \BBA\ Sayers, E.~W. \BBOP2012\BBCP.
\newblock \BBOQ Genbank\BBCQ\
\newblock {\Bem Nucleic Acids Research}, {\Bem 41\/}(D1), D36--D42.

\bibitem[\protect\BCAY{Berg, Tymoczko,\ \BBA\ Stryer}{Berg
  et~al.}{2002}]{berg_biochemistry_2002}
Berg, J.~M., Tymoczko, J.~L., \BBA\ Stryer, L. \BBOP2002\BBCP.
\newblock {\Bem Biochemistry\/} (5 \BEd).
\newblock W H Freeman, New York.

\bibitem[\protect\BCAY{Breiman}{Breiman}{2001}]{breiman_random_2001}
Breiman, L. \BBOP2001\BBCP.
\newblock \BBOQ Random forests\BBCQ\
\newblock {\Bem Machine learning}, {\Bem 45\/}(1), 5--32.

\bibitem[\protect\BCAY{Bukhari, Bukhari, Syed,\ \BBA\ Shah}{Bukhari
  et~al.}{2020}]{bukhari2020diagnostic}
Bukhari, S. U.~K., Bukhari, S. S.~K., Syed, A., \BBA\ Shah, S. S.~H.
  \BBOP2020\BBCP.
\newblock \BBOQ The diagnostic evaluation of convolutional neural network
  ({CNN}) for the assessment of chest {X}-ray of patients infected with
  {COVID}-19\BBCQ\
\newblock {\Bem medRxiv preprint medRxiv:2020.03.26.20044610v1}.

\bibitem[\protect\BCAY{Bung, Krishnan, Bulusu,\ \BBA\ Roy}{Bung
  et~al.}{2020}]{bung_novo_2020}
Bung, N., Krishnan, S.~R., Bulusu, G., \BBA\ Roy, A. \BBOP2020\BBCP.
\newblock \BBOQ De novo design of new chemical entities ({{NCEs}}) for
  {{SARS}}-{{CoV}}-2 using {Artificial Intelligence}\BBCQ\
\newblock {\Bem ChemRxiv preprint chemRxiv:11998347.v2}.

\bibitem[\protect\BCAY{Burley, Berman, Kleywegt, Markley, Nakamura,\ \BBA\
  Velankar}{Burley et~al.}{2017}]{burley2017protein}
Burley, S.~K., Berman, H.~M., Kleywegt, G.~J., Markley, J.~L., Nakamura, H.,
  \BBA\ Velankar, S. \BBOP2017\BBCP.
\newblock \BBOQ Protein {D}ata {B}ank ({PDB}): {T}he single global
  macromolecular structure archive\BBCQ\
\newblock In {\Bem Protein Crystallography}, \BPGS\ 627--641. Springer.

\bibitem[\protect\BCAY{Carrillo-Larco\ \BBA\ Castillo-Cara}{Carrillo-Larco\
  \BBA\ Castillo-Cara}{2020}]{larco2020using}
Carrillo-Larco, R.\BBACOMMA\  \BBA\ Castillo-Cara, M. \BBOP2020\BBCP.
\newblock \BBOQ Using country-level variables to classify countries according
  to the number of confirmed {COVID}-19 cases: An unsupervised machine learning
  approach\BBCQ\
\newblock {\Bem Wellcome Open Research}, {\Bem 5\/}(56).

\bibitem[\protect\BCAY{{CAS}}{{CAS}}{2020}]{CAS}
{CAS} \BBOP2020\BBCP.
\newblock \BBOQ {COVID}-19 antiviral candidate compounds dataset\BBCQ\
\newblock
  \urllink{https://www.cas.org/covid-19-antiviral-compounds-dataset}{https://www.cas.org/
  covid-19-antiviral-compounds-dataset}.

\bibitem[\protect\BCAY{Cascella, Rajnik, Cuomo, Dulebohn,\ \BBA\
  Di~Napoli}{Cascella et~al.}{2020}]{cascella2020features}
Cascella, M., Rajnik, M., Cuomo, A., Dulebohn, S.~C., \BBA\ Di~Napoli, R.
  \BBOP2020\BBCP.
\newblock \BBOQ Features, evaluation and treatment coronavirus
  ({COVID}-19)\BBCQ\
\newblock In {\Bem StatPearls [Internet]}. StatPearls Publishing.

\bibitem[\protect\BCAY{CDC}{CDC}{2020}]{cdcfaq}
CDC \BBOP2020\BBCP.
\newblock \BBOQ Coronavirus disease 2019 ({COVID}-19) - {F}requently asked
  questions\BBCQ\
\newblock \urllink{https://www.cdc.gov/coronavirus/2019-ncov/faq.html}{https://
  www.cdc.gov/coronavirus/2019-ncov/faq.html}.

\bibitem[\protect\BCAY{Chen, Khodadoust, Olsson, Wagar, Fast, Liu, Muftuoglu,
  Sworder, Diehn,\ \BBA\ Levy}{Chen et~al.}{2019}]{chen_predicting_2019}
Chen, B., Khodadoust, M.~S., Olsson, N., Wagar, L.~E., Fast, E., Liu, C.~L.,
  Muftuoglu, Y., Sworder, B.~J., Diehn, M., \BBA\ Levy, R. \BBOP2019\BBCP.
\newblock \BBOQ Predicting {{HLA}} class {{II}} antigen presentation through
  integrated deep learning\BBCQ\
\newblock {\Bem Nature Biotechnology}, {\Bem 37\/}(11), 1332--1343.

\bibitem[\protect\BCAY{Chen, Lerman,\ \BBA\ Ferrara}{Chen
  et~al.}{2020a}]{chen2020covid19}
Chen, E., Lerman, K., \BBA\ Ferrara, E. \BBOP2020a\BBCP.
\newblock \BBOQ {COVID}-19: The first public coronavirus {Twitter}
  dataset\BBCQ\
\newblock {\Bem arXiv preprint arXiv:2003.07372}.

\bibitem[\protect\BCAY{Chen, Lerman,\ \BBA\ Ferrara}{Chen
  et~al.}{2020b}]{chen2020tracking}
Chen, E., Lerman, K., \BBA\ Ferrara, E. \BBOP2020b\BBCP.
\newblock \BBOQ Tracking social media discourse about the {COVID}-19 pandemic:
  {D}evelopment of a public coronavirus {Twitter} data set\BBCQ\
\newblock {\Bem JMIR Public Health and Surveillance}, {\Bem 6\/}(2), e19273.

\bibitem[\protect\BCAY{Chen, Wu, Zhang, Zhang, Gong, Zhao, Hu, Wang, Hu, Zheng,
  et~al.}{Chen et~al.}{2020c}]{chen2020deep}
Chen, J., Wu, L., Zhang, J., Zhang, L., Gong, D., Zhao, Y., Hu, S., Wang, Y.,
  Hu, X., Zheng, B., et~al. \BBOP2020c\BBCP.
\newblock \BBOQ Deep learning-based model for detecting 2019 novel coronavirus
  pneumonia on high-resolution computed tomography: {A} prospective study\BBCQ\
\newblock {\Bem medRxiv preprint medRxiv:2020.02.25.20021568}.

\bibitem[\protect\BCAY{Chen, Allot,\ \BBA\ Lu}{Chen
  et~al.}{2020d}]{chen2020keep}
Chen, Q., Allot, A., \BBA\ Lu, Z. \BBOP2020d\BBCP.
\newblock \BBOQ Keep up with the latest coronavirus research\BBCQ\
\newblock {\Bem Nature}, {\Bem 579\/}(7798), 193--193.

\bibitem[\protect\BCAY{Chen\ \BBA\ Guestrin}{Chen\ \BBA\
  Guestrin}{2016}]{chen2016xgboost}
Chen, T.\BBACOMMA\  \BBA\ Guestrin, C. \BBOP2016\BBCP.
\newblock \BBOQ Xgboost: A scalable tree boosting system\BBCQ\
\newblock In {\Bem Proceedings of the 22nd {ACM} {SIGKDD} International
  Conference on Knowledge Discovery and Data Mining}, \BPGS\ 785--794.

\bibitem[\protect\BCAY{Chenthamarakshan, Das, Padhi, Strobelt, Lim, Hoover,
  Hoffman,\ \BBA\ Mojsilovic}{Chenthamarakshan
  et~al.}{2020}]{chenthamarakshan_target-specific_2020}
Chenthamarakshan, V., Das, P., Padhi, I., Strobelt, H., Lim, K.~W., Hoover, B.,
  Hoffman, S.~C., \BBA\ Mojsilovic, A. \BBOP2020\BBCP.
\newblock \BBOQ Target-specific and selective drug design for {{COVID}}-19
  using deep generative models\BBCQ\
\newblock {\Bem arXiv preprint arXiv:2004.01215}.

\bibitem[\protect\BCAY{Cho, Van~Merri{\"e}nboer, Gulcehre, Bahdanau, Bougares,
  Schwenk,\ \BBA\ Bengio}{Cho et~al.}{2014}]{cho2014learning}
Cho, K., Van~Merri{\"e}nboer, B., Gulcehre, C., Bahdanau, D., Bougares, F.,
  Schwenk, H., \BBA\ Bengio, Y. \BBOP2014\BBCP.
\newblock \BBOQ Learning phrase representations using {RNN} encoder-decoder for
  statistical machine translation\BBCQ\
\newblock {\Bem arXiv preprint arXiv:1406.1078}.

\bibitem[\protect\BCAY{Cinelli, Quattrociocchi, Galeazzi, Valensise, Brugnoli,
  Schmidt, Zola, Zollo,\ \BBA\ Scala}{Cinelli
  et~al.}{2020}]{cinelli2020infodemic}
Cinelli, M., Quattrociocchi, W., Galeazzi, A., Valensise, C.~M., Brugnoli, E.,
  Schmidt, A.~L., Zola, P., Zollo, F., \BBA\ Scala, A. \BBOP2020\BBCP.
\newblock \BBOQ The {COVID}-19 social media infodemic\BBCQ\
\newblock {\Bem arXiv preprint arXiv:2003.05004}.

\bibitem[\protect\BCAY{Cohen, Morrison,\ \BBA\ Dao}{Cohen
  et~al.}{2020}]{cohen2020covid}
Cohen, J.~P., Morrison, P., \BBA\ Dao, L. \BBOP2020\BBCP.
\newblock \BBOQ {COVID}-19 image data collection\BBCQ\
\newblock {\Bem arXiv preprint arXiv:2003.11597}.

\bibitem[\protect\BCAY{Cortes\ \BBA\ Vapnik}{Cortes\ \BBA\
  Vapnik}{1995}]{cortes1995support}
Cortes, C.\BBACOMMA\  \BBA\ Vapnik, V. \BBOP1995\BBCP.
\newblock \BBOQ Support-vector networks\BBCQ\
\newblock {\Bem Machine learning}, {\Bem 20\/}(3), 273--297.

\bibitem[\protect\BCAY{Corum\ \BBA\ Zimmer}{Corum\ \BBA\
  Zimmer}{2020a}]{corum_bad_2020}
Corum, J.\BBACOMMA\  \BBA\ Zimmer, C. \BBOP2020a\BBCP.
\newblock \BBOQ Bad news wrapped in protein: {I}nside the coronavirus
  genome\BBCQ\
\newblock {\Bem The New York Times}.

\bibitem[\protect\BCAY{Corum\ \BBA\ Zimmer}{Corum\ \BBA\
  Zimmer}{2020b}]{corum_how_2020}
Corum, J.\BBACOMMA\  \BBA\ Zimmer, C. \BBOP2020b\BBCP.
\newblock \BBOQ How coronavirus hijacks your cells\BBCQ\
\newblock {\Bem The New York Times}.

\bibitem[\protect\BCAY{{COVID}}{{COVID}}{2020}]{clearinghouse}
{COVID}, U.~A. \BBOP2020\BBCP.
\newblock \BBOQ {COVID}-19 dataset clearinghouse\BBCQ\
\newblock
  \urllink{https://discourse.data-against-covid.org/c/i-have-data/15}{https://discourse.data-against-covid.org/c/i-have-data/15}.

\bibitem[\protect\BCAY{Dandekar\ \BBA\ Barbastathis}{Dandekar\ \BBA\
  Barbastathis}{2020}]{dandekar2020neural}
Dandekar, R.\BBACOMMA\  \BBA\ Barbastathis, G. \BBOP2020\BBCP.
\newblock \BBOQ Neural network aided quarantine control model estimation of
  covid spread in {W}uhan, {C}hina\BBCQ\
\newblock {\Bem arXiv preprint arXiv:2003.09403}.

\bibitem[\protect\BCAY{Davies, Nowotka, Papadatos, Dedman, Gaulton, Atkinson,
  Bellis,\ \BBA\ Overington}{Davies et~al.}{2015}]{davies2015chembl}
Davies, M., Nowotka, M., Papadatos, G., Dedman, N., Gaulton, A., Atkinson, F.,
  Bellis, L., \BBA\ Overington, J.~P. \BBOP2015\BBCP.
\newblock \BBOQ {ChEMBL} web services: Streamlining access to drug discovery
  data and utilities\BBCQ\
\newblock {\Bem Nucleic Acids Research}, {\Bem 43\/}(W1), W612--W620.

\bibitem[\protect\BCAY{Deng, Dong, Socher, Li, Li,\ \BBA\ Fei-Fei}{Deng
  et~al.}{2009}]{deng2009imagenet}
Deng, J., Dong, W., Socher, R., Li, L.-J., Li, K., \BBA\ Fei-Fei, L.
  \BBOP2009\BBCP.
\newblock \BBOQ Imagenet: A large-scale hierarchical image database\BBCQ\
\newblock In {\Bem 2009 IEEE conference on computer vision and pattern
  recognition}, \BPGS\ 248--255. IEEE.

\bibitem[\protect\BCAY{Devlin, Chang, Lee,\ \BBA\ Toutanova}{Devlin
  et~al.}{2019}]{devlin_bert_2019}
Devlin, J., Chang, M.-W., Lee, K., \BBA\ Toutanova, K. \BBOP2019\BBCP.
\newblock \BBOQ {BERT}: {{Pre}}-training of deep bidirectional transformers for
  language understanding\BBCQ\
\newblock In {\Bem Proceedings of {NAACL-HLT} 2019}, \BPGS\ 4171--4186.
  Association for Computational Linguistics.

\bibitem[\protect\BCAY{{Dimensions AI}}{{Dimensions AI}}{2020}]{dimensions}
{Dimensions AI} \BBOP2020\BBCP.
\newblock \BBOQ Dimensions {COVID}-19 publications, datasets and clinical
  trials\BBCQ\
\newblock
  \urllink{https://covid-19.dimensions.ai/}{https://covid-19.dimensions.ai/}.

\bibitem[\protect\BCAY{Dong, Du,\ \BBA\ Gardner}{Dong
  et~al.}{2020}]{dong2020interactive}
Dong, E., Du, H., \BBA\ Gardner, L. \BBOP2020\BBCP.
\newblock \BBOQ An interactive web-based dashboard to track {COVID}-19 in real
  time\BBCQ\
\newblock {\Bem The Lancet infectious diseases}, {\Bem 20\/}(5), 533--534.

\bibitem[\protect\BCAY{Donner, Kazmierczak,\ \BBA\ Fortney}{Donner
  et~al.}{2018}]{donner_drug_2018}
Donner, Y., Kazmierczak, S., \BBA\ Fortney, K. \BBOP2018\BBCP.
\newblock \BBOQ Drug repurposing using deep embeddings of gene expression
  profiles\BBCQ\
\newblock {\Bem Molecular Pharmaceutics}, {\Bem 15\/}(10), 4314--4325.

\bibitem[\protect\BCAY{Dror\ \BBA\ Huang}{Dror\ \BBA\
  Huang}{2019}]{dror_cs279_2019}
Dror, R.\BBACOMMA\  \BBA\ Huang, P. \BBOP2019\BBCP.
\newblock \BBOQ {CS279} computational biology: {S}tructure and organization of
  biomolecules and cells\BBCQ\
\newblock
  \urllink{https://web.stanford.edu/class/cs279/}{https://web.stanford.edu/class/cs279/}.

\bibitem[\protect\BCAY{Fang, Zhang, Xie, Lin, Ying, Pang,\ \BBA\ Ji}{Fang
  et~al.}{2020}]{fang2020sensitivity}
Fang, Y., Zhang, H., Xie, J., Lin, M., Ying, L., Pang, P., \BBA\ Ji, W.
  \BBOP2020\BBCP.
\newblock \BBOQ Sensitivity of chest {CT} for {COVID}-19: {C}omparison to
  {RT-PCR}\BBCQ\
\newblock {\Bem Radiology}, 200432.

\bibitem[\protect\BCAY{Fast, Altman,\ \BBA\ Chen}{Fast
  et~al.}{2020}]{fast_potential_2020}
Fast, E., Altman, R.~B., \BBA\ Chen, B. \BBOP2020\BBCP.
\newblock \BBOQ Potential {{T}}-cell and {{B}}-cell epitopes of
  2019-{{nCoV}}\BBCQ\
\newblock {\Bem bioRxiv preprint bioRxiv:2020.02.19.955484v2}.

\bibitem[\protect\BCAY{Fauqueur, Thillaisundaram,\ \BBA\ Togia}{Fauqueur
  et~al.}{2019}]{fauqueur_constructing_2019}
Fauqueur, J., Thillaisundaram, A., \BBA\ Togia, T. \BBOP2019\BBCP.
\newblock \BBOQ Constructing large scale biomedical knowledge bases from
  scratch with rapid annotation of interpretable patterns\BBCQ\
\newblock In {\Bem Proceedings of the {BioNLP} 2019 Workshop}, \BPGS\ 142--151.

\bibitem[\protect\BCAY{Feng, Huang, Wang, Chen, Zhai, Zhu, Chen, Wang, Su,
  Huang, et~al.}{Feng et~al.}{2020}]{feng2020novel}
Feng, C., Huang, Z., Wang, L., Chen, X., Zhai, Y., Zhu, F., Chen, H., Wang, Y.,
  Su, X., Huang, S., et~al. \BBOP2020\BBCP.
\newblock \BBOQ A novel triage tool of artificial intelligence assisted
  diagnosis aid system for suspected {COVID}-19 pneumonia in fever
  clinics\BBCQ\
\newblock {\Bem medRxiv preprint medRxiv:2020.03.19.20039099v1}.

\bibitem[\protect\BCAY{Fisman, Hauck, Tuite,\ \BBA\ Greer}{Fisman
  et~al.}{2013}]{fisman2013idea}
Fisman, D.~N., Hauck, T.~S., Tuite, A.~R., \BBA\ Greer, A.~L. \BBOP2013\BBCP.
\newblock \BBOQ An {IDEA} for short term outbreak projection: {N}earcasting
  using the basic reproduction number\BBCQ\
\newblock {\Bem PLOS One}, {\Bem 8\/}(12).

\bibitem[\protect\BCAY{Fister, Fister,\ \BBA\ Fister}{Fister
  et~al.}{2020}]{fister2020discovering}
Fister, I.~J., Fister, K., \BBA\ Fister, I. \BBOP2020\BBCP.
\newblock \BBOQ Discovering associations in {COVID}-19 related research
  papers\BBCQ\
\newblock {\Bem arXiv preprint arXiv:2004.00673}.

\bibitem[\protect\BCAY{Fong, Li, Dey, Crespo,\ \BBA\ Herrera-Viedma}{Fong
  et~al.}{2020}]{fong2020finding}
Fong, S.~J., Li, G., Dey, N., Crespo, R.~G., \BBA\ Herrera-Viedma, E.
  \BBOP2020\BBCP.
\newblock \BBOQ Finding an accurate early forecasting model from small dataset:
  A case of 2019-ncov novel coronavirus outbreak\BBCQ\
\newblock {\Bem International Journal of Interactive Multimedia and Artificial
  Intelligence}, {\Bem 6\/}(1), 132--140.

\bibitem[\protect\BCAY{Gallotti, Valle, Castaldo, Sacco,\ \BBA\
  Domenico}{Gallotti et~al.}{2020}]{gallotti2020assessing}
Gallotti, R., Valle, F., Castaldo, N., Sacco, P., \BBA\ Domenico, M.~D.
  \BBOP2020\BBCP.
\newblock \BBOQ Assessing the risks of ``infodemics” in response to
  {COVID}-19 epidemics\BBCQ\
\newblock {\Bem arXiv preprint arXiv:2004.03997}.

\bibitem[\protect\BCAY{{GDELT Project}}{{GDELT Project}}{2020}]{tv}
{GDELT Project} \BBOP2020\BBCP.
\newblock \BBOQ Coronavirus ({COVID}-19) television coverage\BBCQ\
\newblock \urllink{https://www.gdeltproject.org}{https://www.gdelt
  project.org/}.

\bibitem[\protect\BCAY{Ge, Tian, Huang, Wan, Li, Li, Yang, Hong, Wu,\ \BBA\
  Yuan}{Ge et~al.}{2020}]{ge_data-driven_2020}
Ge, Y., Tian, T., Huang, S., Wan, F., Li, J., Li, S., Yang, H., Hong, L., Wu,
  N., \BBA\ Yuan, E. \BBOP2020\BBCP.
\newblock \BBOQ A data-driven drug repositioning framework discovered a
  potential therapeutic agent targeting {{COVID}}-19\BBCQ\
\newblock {\Bem bioRxiv preprint bioRxiv:2020.03.11.986836}.

\bibitem[\protect\BCAY{{GHDDI}}{{GHDDI}}{2020}]{GHDDI}
{GHDDI} \BBOP2020\BBCP.
\newblock \BBOQ Targeting {COVID}-19: {GHDDI} info sharing portal\BBCQ\
\newblock
  \urllink{https://ghddi-ailab.github.io/Targeting2019-nCoV/}{https://ghddi-ailab.github.io/Targeting2019-nCoV/}.

\bibitem[\protect\BCAY{Ghoshal\ \BBA\ Tucker}{Ghoshal\ \BBA\
  Tucker}{2020}]{ghoshal2020estimating}
Ghoshal, B.\BBACOMMA\  \BBA\ Tucker, A. \BBOP2020\BBCP.
\newblock \BBOQ Estimating uncertainty and interpretability in deep learning
  for coronavirus ({COVID}-19) detection\BBCQ\
\newblock {\Bem arXiv preprint arXiv:2003.10769}.

\bibitem[\protect\BCAY{Goodfellow, {Pouget-Abadie}, Mirza, Xu, {Warde-Farley},
  Ozair, Courville,\ \BBA\ Bengio}{Goodfellow
  et~al.}{2014}]{goodfellow_generative_2014-1}
Goodfellow, I., {Pouget-Abadie}, J., Mirza, M., Xu, B., {Warde-Farley}, D.,
  Ozair, S., Courville, A., \BBA\ Bengio, Y. \BBOP2014\BBCP.
\newblock \BBOQ Generative adversarial nets\BBCQ\
\newblock In {\Bem Advances in Neural Information Processing Systems}, \BPGS\
  2672--2680.

\bibitem[\protect\BCAY{Gozes, Frid-Adar, Greenspan, Browning, Zhang, Ji,
  Bernheim,\ \BBA\ Siegel}{Gozes et~al.}{2020a}]{gozes2020rapid}
Gozes, O., Frid-Adar, M., Greenspan, H., Browning, P.~D., Zhang, H., Ji, W.,
  Bernheim, A., \BBA\ Siegel, E. \BBOP2020a\BBCP.
\newblock \BBOQ Rapid {AI} development cycle for the coronavirus ({COVID}-19)
  pandemic: Initial results for automated detection \& patient monitoring using
  deep learning {CT} image analysis\BBCQ\
\newblock {\Bem arXiv preprint arXiv:2003.05037}.

\bibitem[\protect\BCAY{Gozes, Frid-Adar, Sagie, Zhang, Ji,\ \BBA\
  Greenspan}{Gozes et~al.}{2020b}]{gozes2020coronavirus}
Gozes, O., Frid-Adar, M., Sagie, N., Zhang, H., Ji, W., \BBA\ Greenspan, H.
  \BBOP2020b\BBCP.
\newblock \BBOQ Coronavirus detection and analysis on chest {CT} with deep
  learning\BBCQ\
\newblock {\Bem arXiv preprint arXiv:2004.02640}.

\bibitem[\protect\BCAY{Gussow, Auslander, Wolf,\ \BBA\ Koonin}{Gussow
  et~al.}{2020}]{gussow_genomic_2020}
Gussow, A.~B., Auslander, N., Wolf, Y.~I., \BBA\ Koonin, E.~V. \BBOP2020\BBCP.
\newblock \BBOQ Genomic determinants of pathogenicity in {{SARS}}-{{CoV}}-2 and
  other human coronaviruses\BBCQ\
\newblock {\Bem Proceedings of the National Academy of Sciences}, {\Bem
  117\/}(26), 15193--15199.

\bibitem[\protect\BCAY{Hadfield, Megill, Bell, Huddleston, Potter, Callender,
  Sagulenko, Bedford,\ \BBA\ Neher}{Hadfield
  et~al.}{2018}]{hadfield2018nextstrain}
Hadfield, J., Megill, C., Bell, S.~M., Huddleston, J., Potter, B., Callender,
  C., Sagulenko, P., Bedford, T., \BBA\ Neher, R.~A. \BBOP2018\BBCP.
\newblock \BBOQ Nextstrain: {R}eal-time tracking of pathogen evolution\BBCQ\
\newblock {\Bem Bioinformatics}, {\Bem 34\/}(23), 4121--4123.

\bibitem[\protect\BCAY{Hammoudi, Benhabiles, Melkemi, Dornaika,
  Arganda-Carreras, Collard,\ \BBA\ Scherpereel}{Hammoudi
  et~al.}{2020}]{hammoudi2020deep}
Hammoudi, K., Benhabiles, H., Melkemi, M., Dornaika, F., Arganda-Carreras, I.,
  Collard, D., \BBA\ Scherpereel, A. \BBOP2020\BBCP.
\newblock \BBOQ Deep learning on chest {X}-ray images to detect and evaluate
  pneumonia cases at the era of {COVID}-19\BBCQ\
\newblock {\Bem arXiv preprint arXiv:2004.03399}.

\bibitem[\protect\BCAY{Hartono}{Hartono}{2019}]{hartono2019mixing}
Hartono, P. \BBOP2019\BBCP.
\newblock \BBOQ Mixing autoencoder with classifier: {C}onceptual data
  visualization\BBCQ\
\newblock {\Bem arXiv preprint arXiv:1912.01137}.

\bibitem[\protect\BCAY{Hartono}{Hartono}{2020}]{hartono2020generating}
Hartono, P. \BBOP2020\BBCP.
\newblock \BBOQ Generating similarity map for {COVID}-19 transmission dynamics
  with topological autoencoder\BBCQ\
\newblock {\Bem arXiv preprint arXiv:2004.01481}.

\bibitem[\protect\BCAY{He, Zhang, Ren,\ \BBA\ Sun}{He
  et~al.}{2016}]{he2016deep}
He, K., Zhang, X., Ren, S., \BBA\ Sun, J. \BBOP2016\BBCP.
\newblock \BBOQ Deep residual learning for image recognition\BBCQ\
\newblock In {\Bem Proceedings of the IEEE conference on computer vision and
  pattern recognition}, \BPGS\ 770--778.

\bibitem[\protect\BCAY{Heo\ \BBA\ Feig}{Heo\ \BBA\
  Feig}{2020}]{heo_modeling_2020}
Heo, L.\BBACOMMA\  \BBA\ Feig, M. \BBOP2020\BBCP.
\newblock \BBOQ Modeling of {{Severe Acute Respiratory Syndrome Coronavirus}} 2
  ({{SARS}}-{{CoV}}-2) {{Proteins}} by {{Machine Learning}} and
  {{Physics}}-{{Based Refinement}}\BBCQ\
\newblock {\Bem bioRxiv preprint bioRxiv:2020.03.25.008904v1}.

\bibitem[\protect\BCAY{Hochreiter\ \BBA\ Schmidhuber}{Hochreiter\ \BBA\
  Schmidhuber}{1997}]{lstms1997}
Hochreiter, S.\BBACOMMA\  \BBA\ Schmidhuber, J. \BBOP1997\BBCP.
\newblock \BBOQ Long short-term memory\BBCQ\
\newblock {\Bem Neural Computation}, {\Bem 9\/}(8), 1735--1780.

\bibitem[\protect\BCAY{Hoffmann, Kleine-Weber, Schroeder, Kr{\"u}ger, Herrler,
  Erichsen, Schiergens, Herrler, Wu, Nitsche, et~al.}{Hoffmann
  et~al.}{2020}]{hoffmann2020sars}
Hoffmann, M., Kleine-Weber, H., Schroeder, S., Kr{\"u}ger, N., Herrler, T.,
  Erichsen, S., Schiergens, T.~S., Herrler, G., Wu, N.-H., Nitsche, A., et~al.
  \BBOP2020\BBCP.
\newblock \BBOQ {SARS-CoV-2} cell entry depends on {ACE2} and {TMPRSS2} and is
  blocked by a clinically proven protease inhibitor\BBCQ\
\newblock {\Bem Cell}, {\Bem 181\/}(2), 271--280.

\bibitem[\protect\BCAY{Hofmarcher, Mayr, Rumetshofer, Ruch, Renz, Schimunek,
  Seidl, Vall, Widrich, Hochreiter,\ \BBA\ Klambauer}{Hofmarcher
  et~al.}{2020}]{hofmarcher_large-scale_2020}
Hofmarcher, M., Mayr, A., Rumetshofer, E., Ruch, P., Renz, P., Schimunek, J.,
  Seidl, P., Vall, A., Widrich, M., Hochreiter, S., \BBA\ Klambauer, G.
  \BBOP2020\BBCP.
\newblock \BBOQ Large-scale ligand-based virtual screening for
  {{SARS}}-{{CoV}}-2 inhibitors using deep neural networks\BBCQ\
\newblock {\Bem arXiv preprint arXiv:2004.00979}.

\bibitem[\protect\BCAY{Homan\ \BBA\ Gelman}{Homan\ \BBA\
  Gelman}{2014}]{homan2014nouturn}
Homan, M.~D.\BBACOMMA\  \BBA\ Gelman, A. \BBOP2014\BBCP.
\newblock \BBOQ The {N}o-{U}-{T}urn sampler: Adaptively setting path lengths in
  {H}amiltonian {M}onte {C}arlo\BBCQ\
\newblock {\Bem Journal of Machine Learning Research}, {\Bem 15\/}(1),
  1593–1623.

\bibitem[\protect\BCAY{Hong, Lin, Li, Wan, Yang, Jiang, Zhao,\ \BBA\ Zeng}{Hong
  et~al.}{2020}]{hong2020novel}
Hong, L., Lin, J., Li, S., Wan, F., Yang, H., Jiang, T., Zhao, D., \BBA\ Zeng,
  J. \BBOP2020\BBCP.
\newblock \BBOQ A novel machine learning framework for automated biomedical
  relation extraction from large-scale literature repositories\BBCQ\
\newblock {\Bem Nature Machine Intelligence}, {\Bem 2}, 347--355.

\bibitem[\protect\BCAY{Hu, Jiang,\ \BBA\ Yin}{Hu
  et~al.}{2020a}]{hu_prediction_2020}
Hu, F., Jiang, J., \BBA\ Yin, P. \BBOP2020a\BBCP.
\newblock \BBOQ Prediction of potential commercially inhibitors against
  {{SARS}}-{{CoV}}-2 by multi-task deep model\BBCQ\
\newblock {\Bem arXiv preprint arXiv:2003.00728}.

\bibitem[\protect\BCAY{Hu, Ge, Li, Boerwinkle, Jin,\ \BBA\ Xiong}{Hu
  et~al.}{2020b}]{hu2020forecasting}
Hu, Z., Ge, Q., Li, S., Boerwinkle, E., Jin, L., \BBA\ Xiong, M.
  \BBOP2020b\BBCP.
\newblock \BBOQ Forecasting and evaluating multiple interventions for
  {COVID-19} worldwide\BBCQ\
\newblock {\Bem Frontiers in Artificial Intelligence}, {\Bem 3}, 41.

\bibitem[\protect\BCAY{Hu, Ge, Li, Jin,\ \BBA\ Xiong}{Hu
  et~al.}{2020c}]{hu2020artificial}
Hu, Z., Ge, Q., Li, S., Jin, L., \BBA\ Xiong, M. \BBOP2020c\BBCP.
\newblock \BBOQ Artificial intelligence forecasting of {COVID}-19 in
  {C}hina\BBCQ\
\newblock {\Bem arXiv preprint arXiv:2002.07112}.

\bibitem[\protect\BCAY{Huang, Chen, Ma,\ \BBA\ Kuo}{Huang
  et~al.}{2020}]{huang2020multiple}
Huang, C.-J., Chen, Y.-H., Ma, Y., \BBA\ Kuo, P.-H. \BBOP2020\BBCP.
\newblock \BBOQ Multiple-input deep convolutional neural network model for
  {COVID}-19 forecasting in {China}\BBCQ\
\newblock {\Bem medRxiv preprint medRxiv:2020.03.23.20041608}.

\bibitem[\protect\BCAY{{Humanitarian Data Exchange}}{{Humanitarian Data
  Exchange}}{2020}]{humxchange}
{Humanitarian Data Exchange} \BBOP2020\BBCP.
\newblock \BBOQ {COVID}-19 pandemic -- {H}umanitarian data exchange\BBCQ\
\newblock
  \urllink{https://data.humdata.org/event/covid-19}{https://data.humdata.org/event/covid-19}.
\newblock Accessed: 2020-08-05.

\bibitem[\protect\BCAY{Imran, Posokhova, Qureshi, Masood, Riaz, Ali, John,
  Hussain,\ \BBA\ Nabeel}{Imran et~al.}{2020}]{imran2020ai4covid}
Imran, A., Posokhova, I., Qureshi, H.~N., Masood, U., Riaz, S., Ali, K., John,
  C.~N., Hussain, I., \BBA\ Nabeel, M. \BBOP2020\BBCP.
\newblock \BBOQ {AI4COVID-19}: {AI} enabled preliminary diagnosis for
  {COVID-19} from cough samples via an app\BBCQ\
\newblock {\Bem Informatics in Medicine Unlocked}, 100378.

\bibitem[\protect\BCAY{Ivakhnenko}{Ivakhnenko}{1970}]{ivakhnenko1970heuristic}
Ivakhnenko, A.~G. \BBOP1970\BBCP.
\newblock \BBOQ Heuritic self-organization in problems of engineering
  cybernetics\BBCQ\
\newblock {\Bem Automatica}, {\Bem 6}, 207--219.

\bibitem[\protect\BCAY{Jang}{Jang}{1993}]{jang1993anfis}
Jang, J.-S.~R. \BBOP1993\BBCP.
\newblock \BBOQ {ANFIS}: adaptive-network-based fuzzy inference system\BBCQ\
\newblock {\Bem IEEE Transactions on Systems, Man, and Cybernetics}, {\Bem
  23\/}(3), 665--685.

\bibitem[\protect\BCAY{Jiang, Coffee, Bari, Wang, Jiang, Huang, Shi, Dai, Cai,
  Zhang, et~al.}{Jiang et~al.}{2020}]{jiang2020towards}
Jiang, X., Coffee, M., Bari, A., Wang, J., Jiang, X., Huang, J., Shi, J., Dai,
  J., Cai, J., Zhang, T., et~al. \BBOP2020\BBCP.
\newblock \BBOQ Towards an artificial intelligence framework for data-driven
  prediction of coronavirus clinical severity\BBCQ\
\newblock {\Bem CMC-Computers, Materials \& Continua}, {\Bem 63}, 537--551.

\bibitem[\protect\BCAY{Jumper, Tunyasuvunakool, Kohli, Hassabis,\ \BBA\
  {AlphaFold Team}}{Jumper et~al.}{2020}]{jumper_computational_2020}
Jumper, J., Tunyasuvunakool, K., Kohli, P., Hassabis, D., \BBA\ {AlphaFold
  Team} \BBOP2020\BBCP.
\newblock \BBOQ Computational predictions of protein structures associated with
  {COVID}-19\BBCQ\
\newblock
  \urllink{https://deepmind.com/research/open-source/computational-predictions-of-protein-structures-associated-with-COVID-19}{https://deepmind.com/research/open-source/computational-predictions-of-protein-structures-associated-with-COVID-19}.

\bibitem[\protect\BCAY{Jurtz, Paul, Andreatta, Marcatili, Peters,\ \BBA\
  Nielsen}{Jurtz et~al.}{2017}]{jurtz_netmhcpan-4.0_2017}
Jurtz, V., Paul, S., Andreatta, M., Marcatili, P., Peters, B., \BBA\ Nielsen,
  M. \BBOP2017\BBCP.
\newblock \BBOQ {{NetMHCpan}}-4.0: Improved peptide\textendash{{MHC}} class
  {{I}} interaction predictions integrating eluted ligand and peptide binding
  affinity data\BBCQ\
\newblock {\Bem The Journal of Immunology}, {\Bem 199\/}(9), 3360--3368.

\bibitem[\protect\BCAY{Kanne, Little, Chung, Elicker,\ \BBA\ Ketai}{Kanne
  et~al.}{2020}]{kanne2020essentials}
Kanne, J.~P., Little, B.~P., Chung, J.~H., Elicker, B.~M., \BBA\ Ketai, L.~H.
  \BBOP2020\BBCP.
\newblock \BBOQ Essentials for radiologists on {COVID}-19: an
  update—radiology scientific expert panel\BBCQ\
\newblock {\Bem Radiology}, 200527.

\bibitem[\protect\BCAY{Karim, D{\"o}hmen, Rebholz-Schuhmann, Decker, Cochez,
  Beyan, et~al.}{Karim et~al.}{2020}]{karim2020deepcovidexplainer}
Karim, M., D{\"o}hmen, T., Rebholz-Schuhmann, D., Decker, S., Cochez, M.,
  Beyan, O., et~al. \BBOP2020\BBCP.
\newblock \BBOQ {DeepCOVIDExplainer}: Explainable {COVID}-19 predictions based
  on chest {X}-ray images\BBCQ\
\newblock {\Bem arXiv preprint arXiv:2004.04582}.

\bibitem[\protect\BCAY{Keenan, Jenkins, Jagodnik, Koplev, He, Torre, Wang,
  Dohlman, Silverstein, Lachmann, et~al.}{Keenan
  et~al.}{2018}]{keenan2018library}
Keenan, A.~B., Jenkins, S.~L., Jagodnik, K.~M., Koplev, S., He, E., Torre, D.,
  Wang, Z., Dohlman, A.~B., Silverstein, M.~C., Lachmann, A., et~al.
  \BBOP2018\BBCP.
\newblock \BBOQ The library of integrated network-based cellular signatures
  {NIH} program: {S}ystem-level cataloging of human cells response to
  perturbations\BBCQ\
\newblock {\Bem Cell systems}, {\Bem 6\/}(1), 13--24.

\bibitem[\protect\BCAY{Kleinberg, van~der Vegt,\ \BBA\ Mozes}{Kleinberg
  et~al.}{2020}]{kleinberg2020measuring}
Kleinberg, B., van~der Vegt, I., \BBA\ Mozes, M. \BBOP2020\BBCP.
\newblock \BBOQ Measuring emotions in the {COVID}-19 real world worry
  dataset\BBCQ\
\newblock {\Bem arXiv preprint arXiv:2004.04225}.

\bibitem[\protect\BCAY{Lampos, Moura, Yom-Tov, Edelstein, Majumder, Hamada,
  Rangaka, McKendry, ,\ \BBA\ Cox}{Lampos et~al.}{2020}]{lampos2020tracking}
Lampos, V., Moura, S., Yom-Tov, E., Edelstein, M., Majumder, M., Hamada, Y.,
  Rangaka, M.~X., McKendry, R.~A., , \BBA\ Cox, I.~J. \BBOP2020\BBCP.
\newblock \BBOQ Tracking {COVID}-19 using online search\BBCQ\
\newblock {\Bem arXiv preprint arXiv:2003.08086}.

\bibitem[\protect\BCAY{LeCun, Bengio,\ \BBA\ Hinton}{LeCun
  et~al.}{2015}]{lecun2015deep}
LeCun, Y., Bengio, Y., \BBA\ Hinton, G. \BBOP2015\BBCP.
\newblock \BBOQ Deep learning\BBCQ\
\newblock {\Bem Nature}, {\Bem 521\/}(7553), 436--444.

\bibitem[\protect\BCAY{Li, Qin, Xu, Yin, Wang, Kong, Bai, Lu, Fang, Song, Cao,
  Liu, Wang, Xu, Fang, Zhang, Xia,\ \BBA\ Xia}{Li
  et~al.}{2020a}]{li2020artificial}
Li, L., Qin, L., Xu, Z., Yin, Y., Wang, X., Kong, B., Bai, J., Lu, Y., Fang,
  Z., Song, Q., Cao, K., Liu, D., Wang, G., Xu, Q., Fang, X., Zhang, S., Xia,
  J., \BBA\ Xia, J. \BBOP2020a\BBCP.
\newblock \BBOQ Using artificial intelligence to detect {COVID-19} and
  community-acquired pneumonia based on pulmonary {CT}: {E}valuation of the
  diagnostic accuracy\BBCQ\
\newblock {\Bem Radiology}, {\Bem 296\/}(2), E65--E71.
\newblock PMID: 32191588.

\bibitem[\protect\BCAY{Li, Li,\ \BBA\ Zhu}{Li
  et~al.}{2020b}]{li2020covidmobilexpert}
Li, X., Li, C., \BBA\ Zhu, D. \BBOP2020b\BBCP.
\newblock \BBOQ {COVID}-{M}obile{X}pert: On-device {COVID}-19 screening using
  snapshots of chest {X}-ray\BBCQ\
\newblock {\Bem arXiv preprint arXiv:2004.03042}.

\bibitem[\protect\BCAY{Liu, Zhou, Li, Garner, Watkins, Carter, Smoot, Gregg,
  Daniels,\ \BBA\ Jervey}{Liu et~al.}{2020a}]{liu_research_2020}
Liu, C., Zhou, Q., Li, Y., Garner, L.~V., Watkins, S.~P., Carter, L.~J., Smoot,
  J., Gregg, A.~C., Daniels, A.~D., \BBA\ Jervey, S. \BBOP2020a\BBCP.
\newblock \BBOQ Research and development on therapeutic agents and vaccines for
  {{COVID}}-19 and related human coronavirus diseases\BBCQ\
\newblock {\Bem ACS Central Science}, {\Bem 6}, 315--331.

\bibitem[\protect\BCAY{Liu, Clemente, Poirier, Ding, Chinazzi, David,
  Vespignani,\ \BBA\ Santillana}{Liu et~al.}{2020b}]{liu2020machine}
Liu, D., Clemente, L., Poirier, C., Ding, X., Chinazzi, M., David, J.~T.,
  Vespignani, A., \BBA\ Santillana, M. \BBOP2020b\BBCP.
\newblock \BBOQ A machine learning methodology for real-time forecasting of the
  2019-2020 {COVID}-19 outbreak using internet searches, news alerts, and
  estimates from mechanistic models\BBCQ\
\newblock {\Bem arXiv preprint arXiv:2004.04019}.

\bibitem[\protect\BCAY{Liu, Lin, Wen, Jorissen,\ \BBA\ Gilson}{Liu
  et~al.}{2007}]{liu2007bindingdb}
Liu, T., Lin, Y., Wen, X., Jorissen, R.~N., \BBA\ Gilson, M.~K. \BBOP2007\BBCP.
\newblock \BBOQ {BindingDB}: A web-accessible database of experimentally
  determined protein--ligand binding affinities\BBCQ\
\newblock {\Bem Nucleic Acids Research}, {\Bem 35\/}(suppl\_1), D198--D201.

\bibitem[\protect\BCAY{{Lopez-Rincon}, Tonda, {Mendoza-Maldonado}, Claassen,
  Garssen,\ \BBA\ Kraneveld}{{Lopez-Rincon}
  et~al.}{2020}]{lopez-rincon_accurate_2020}
{Lopez-Rincon}, A., Tonda, A., {Mendoza-Maldonado}, L., Claassen, E., Garssen,
  J., \BBA\ Kraneveld, A.~D. \BBOP2020\BBCP.
\newblock \BBOQ Accurate identification of {SARS-CoV-2} from viral genome
  sequences using deep learning\BBCQ\
\newblock {\Bem bioRxiv preprint bioRxiv:2020.03.13.990242v1}.

\bibitem[\protect\BCAY{Lu, Hattab, Clemente, Biggerstaff,\ \BBA\ Santillana}{Lu
  et~al.}{2019}]{lu2019improved}
Lu, F.~S., Hattab, M.~W., Clemente, C.~L., Biggerstaff, M., \BBA\ Santillana,
  M. \BBOP2019\BBCP.
\newblock \BBOQ Improved state-level influenza nowcasting in the united states
  leveraging internet-based data and network approaches\BBCQ\
\newblock {\Bem Nature Communications}, {\Bem 10\/}(1), 1--10.

\bibitem[\protect\BCAY{Luccioni, Bullock, {Hoffmann Pham}, Lam,\ \BBA\
  Luengo-Oroz}{Luccioni et~al.}{2020}]{luccioni2020considerations}
Luccioni, A., Bullock, J., {Hoffmann Pham}, K., Lam, C. S.~N., \BBA\
  Luengo-Oroz, M. \BBOP2020\BBCP.
\newblock \BBOQ Considerations, good practices, risks and pitfalls in
  developing {AI} solutions against {COVID}-19\BBCQ\
\newblock In {\Bem Harvard {CRCS} Workshop on {AI} for Social Good}.

\bibitem[\protect\BCAY{Luengo-Oroz, Pham, Bullock, Kirkpatrick, Luccioni,
  Rubel, Wachholz, Chakchouk, Biggs, Nguyen, Purnat,\ \BBA\
  Mariano}{Luengo-Oroz et~al.}{2020}]{luengo-oroz2020artificial}
Luengo-Oroz, M., Pham, K.~H., Bullock, J., Kirkpatrick, R., Luccioni, A.,
  Rubel, S., Wachholz, C., Chakchouk, M., Biggs, P., Nguyen, T., Purnat, T.,
  \BBA\ Mariano, B. \BBOP2020\BBCP.
\newblock \BBOQ Artificial intelligence cooperation to support the global
  response to {COVID}-19\BBCQ\
\newblock {\Bem Nature Machine Intelligence}, {\Bem 2}, 295--297.

\bibitem[\protect\BCAY{Luengo-Oroz, Ledesma-Carbayo, Peyri{\'e}ras,\ \BBA\
  Santos}{Luengo-Oroz et~al.}{2011}]{luengo2011image}
Luengo-Oroz, M.~A., Ledesma-Carbayo, M.~J., Peyri{\'e}ras, N., \BBA\ Santos, A.
  \BBOP2011\BBCP.
\newblock \BBOQ Image analysis for understanding embryo development: {A} bridge
  from microscopy to biological insights\BBCQ\
\newblock {\Bem Current Opinion in Genetics \& Development}, {\Bem 21\/}(5),
  630--637.

\bibitem[\protect\BCAY{Magar, Yadav,\ \BBA\ Farimani}{Magar
  et~al.}{2020}]{magar_potential_2020}
Magar, R., Yadav, P., \BBA\ Farimani, A.~B. \BBOP2020\BBCP.
\newblock \BBOQ Potential neutralizing antibodies discovered for novel
  coronavirus using {M}achine {L}earning\BBCQ\
\newblock {\Bem arXiv preprint arXiv:2003.08447}.

\bibitem[\protect\BCAY{Maghdid, Ghafoor, Sadiq, Curran,\ \BBA\ Rabie}{Maghdid
  et~al.}{2020}]{maghdid2020novel}
Maghdid, H.~S., Ghafoor, K.~Z., Sadiq, A.~S., Curran, K., \BBA\ Rabie, K.
  \BBOP2020\BBCP.
\newblock \BBOQ A novel {AI}-enabled framework to diagnose coronavirus
  {COVID}-19 using smartphone embedded sensors: Design study\BBCQ\
\newblock {\Bem arXiv preprint arXiv:2003.07434}.

\bibitem[\protect\BCAY{Makhzani, Shlens, Jaitly, Goodfellow,\ \BBA\
  Frey}{Makhzani et~al.}{2016}]{makhzani_adversarial_2016}
Makhzani, A., Shlens, J., Jaitly, N., Goodfellow, I., \BBA\ Frey, B.
  \BBOP2016\BBCP.
\newblock \BBOQ Adversarial autoencoders\BBCQ\
\newblock In {\Bem International Conference on Learning Representations}.

\bibitem[\protect\BCAY{Mejova\ \BBA\ Kalimeri}{Mejova\ \BBA\
  Kalimeri}{2020}]{mejova2020advertisers}
Mejova, Y.\BBACOMMA\  \BBA\ Kalimeri, K. \BBOP2020\BBCP.
\newblock \BBOQ Advertisers jump on coronavirus bandwagon: Politics, news, and
  business\BBCQ\
\newblock {\Bem arXiv preprint arXiv:2003.00923}.

\bibitem[\protect\BCAY{Mendez, Gaulton, Bento, Chambers, De~Veij, F{\'e}lix,
  Magari{\~n}os, Mosquera, Mutowo, Nowotka, et~al.}{Mendez
  et~al.}{2019}]{mendez2019chembl}
Mendez, D., Gaulton, A., Bento, A.~P., Chambers, J., De~Veij, M., F{\'e}lix,
  E., Magari{\~n}os, M.~P., Mosquera, J.~F., Mutowo, P., Nowotka, M., et~al.
  \BBOP2019\BBCP.
\newblock \BBOQ {ChEMBL}: Towards direct deposition of bioassay data\BBCQ\
\newblock {\Bem Nucleic Acids Research}, {\Bem 47\/}(D1), D930--D940.

\bibitem[\protect\BCAY{Metsky, Freije, Kosoko-Thoroddsen, Sabeti,\ \BBA\
  Myhrvold}{Metsky et~al.}{2020}]{metsky2020CRISPR}
Metsky, H.~C., Freije, C.~A., Kosoko-Thoroddsen, T.-S.~F., Sabeti, P.~C., \BBA\
  Myhrvold, C. \BBOP2020\BBCP.
\newblock \BBOQ {CRISPR}-based surveillance for {COVID}-19 using
  genomically-comprehensive machine learning design\BBCQ\
\newblock {\Bem bioRxiv preprint bioRxiv:2020.02.26.967026}.

\bibitem[\protect\BCAY{Mezei}{Mezei}{2020}]{archive}
Mezei, K. \BBOP2020\BBCP.
\newblock \BBOQ Archived {COVID}-19 related, news, academic articles,
  essays\BBCQ\
\newblock \urllink{https://www.covid19-archive.com/}{https://
  www.covid19-archive.com/}.

\bibitem[\protect\BCAY{Milletari, Navab,\ \BBA\ Ahmadi}{Milletari
  et~al.}{2016}]{milletari2016}
Milletari, F., Navab, N., \BBA\ Ahmadi, S.-A. \BBOP2016\BBCP.
\newblock \BBOQ V-net: Fully convolutional neural networks for volumetric
  medical image segmentation\BBCQ\
\newblock In {\Bem 2016 Fourth International Conference on 3D Vision (3DV)},
  \BPGS\ 565--571. IEEE.

\bibitem[\protect\BCAY{Mirjalili, Gandomi, Mirjalili, Saremi,\ \BBA\
  Faris}{Mirjalili et~al.}{2017}]{mirjalili2017salpsa}
Mirjalili, S.~M., Gandomi, A.~H., Mirjalili, S.~Z., Saremi, S., \BBA\ Faris, H.
  \BBOP2017\BBCP.
\newblock \BBOQ Salp swarm algorithm: A bio-inspired optimizer for engineering
  design problems\BBCQ\
\newblock {\Bem Advances in Engineering Software}, {\Bem 114}, 163--191.

\bibitem[\protect\BCAY{Mirza\ \BBA\ Osindero}{Mirza\ \BBA\
  Osindero}{2014}]{mirza2014conditional}
Mirza, M.\BBACOMMA\  \BBA\ Osindero, S. \BBOP2014\BBCP.
\newblock \BBOQ Conditional generative adversarial nets\BBCQ\
\newblock {\Bem arXiv preprint arXiv:1411.1784}.

\bibitem[\protect\BCAY{Mizumoto, Kagaya, Zarebski,\ \BBA\ Chowell}{Mizumoto
  et~al.}{2020}]{mizumoto2020asymptomatic}
Mizumoto, K., Kagaya, K., Zarebski, A., \BBA\ Chowell, G. \BBOP2020\BBCP.
\newblock \BBOQ Estimating the asymptomatic proportion of coronavirus disease
  2019 ({COVID}-19) cases on board the {D}iamond {P}rincess cruise ship,
  {Y}okohama, {J}apan, 2020\BBCQ\
\newblock {\Bem Eurosurveillance}, {\Bem 25\/}(10).

\bibitem[\protect\BCAY{Murphy, Smits, Knoops, Korst, Samson, Scholten,
  Schalekamp, Schaefer-Prokop, Philipsen, Meijers, et~al.}{Murphy
  et~al.}{2020}]{murphy2020covid}
Murphy, K., Smits, H., Knoops, A.~J., Korst, M.~B., Samson, T., Scholten,
  E.~T., Schalekamp, S., Schaefer-Prokop, C.~M., Philipsen, R.~H., Meijers, A.,
  et~al. \BBOP2020\BBCP.
\newblock \BBOQ {COVID-19} on the chest radiograph: A multi-reader evaluation
  of an {AI} system\BBCQ\
\newblock {\Bem Radiology}, 201874.

\bibitem[\protect\BCAY{Nagendran, Chen, Lovejoy, Gordon, Komorowski, Harvey,
  Topol, Ioannidis, Collins,\ \BBA\ Maruthappu}{Nagendran
  et~al.}{2020}]{nagendran2020artificial}
Nagendran, M., Chen, Y., Lovejoy, C.~A., Gordon, A.~C., Komorowski, M., Harvey,
  H., Topol, E.~J., Ioannidis, J.~P., Collins, G.~S., \BBA\ Maruthappu, M.
  \BBOP2020\BBCP.
\newblock \BBOQ Artificial intelligence versus clinicians: {S}ystematic review
  of design, reporting standards, and claims of deep learning studies\BBCQ\
\newblock {\Bem BMJ}, {\Bem 368}.

\bibitem[\protect\BCAY{{Nexoid}}{{Nexoid}}{2020}]{calculator}
{Nexoid} \BBOP2020\BBCP.
\newblock \BBOQ {COVID}-19 (coronavirus) survival calculator\BBCQ\
\newblock
  \urllink{https://www.covid19survivalcalculator.com/calculator}{https://www.covid19survival
  calculator.com/calculator}.

\bibitem[\protect\BCAY{Ng, Lee, Yang, Yang, Li, Wang, Lui, Lo, Leung, Khong,
  et~al.}{Ng et~al.}{2020}]{ng2020imaging}
Ng, M.-Y., Lee, E.~Y., Yang, J., Yang, F., Li, X., Wang, H., Lui, M. M.-s., Lo,
  C. S.-Y., Leung, B., Khong, P.-L., et~al. \BBOP2020\BBCP.
\newblock \BBOQ Imaging profile of the {COVID}-19 infection: {R}adiologic
  findings and literature review\BBCQ\
\newblock {\Bem Radiology: Cardiothoracic Imaging}, {\Bem 2\/}(1), e200034.

\bibitem[\protect\BCAY{Nguyen, Gao, Chen, Wang,\ \BBA\ Wei}{Nguyen
  et~al.}{2020a}]{nguyen_potentially_2020}
Nguyen, D.~D., Gao, K., Chen, J., Wang, R., \BBA\ Wei, G. \BBOP2020a\BBCP.
\newblock \BBOQ Potentially highly potent drugs for 2019-{nCoV}\BBCQ\
\newblock {\Bem bioRxiv preprint bioRxiv:2020.02.05.936013v1}.

\bibitem[\protect\BCAY{Nguyen, Gao, Wang,\ \BBA\ Wei}{Nguyen
  et~al.}{2020b}]{nguyen_machine_2020}
Nguyen, D.~D., Gao, K., Wang, R., \BBA\ Wei, G. \BBOP2020b\BBCP.
\newblock \BBOQ Machine intelligence design of 2019-{{nCoV}} drugs\BBCQ\
\newblock {\Bem bioRxiv preprint bioRxiv:2020.01.30.927889}.

\bibitem[\protect\BCAY{Ong, Wong, Huffman,\ \BBA\ He}{Ong
  et~al.}{2020}]{ong_covid-19_2020}
Ong, E., Wong, M.~U., Huffman, A., \BBA\ He, Y. \BBOP2020\BBCP.
\newblock \BBOQ {{COVID}}-19 coronavirus vaccine design using reverse
  vaccinology and machine learning\BBCQ\
\newblock {\Bem Frontiers in Immunology}, {\Bem 11\/}(1581), 1--13.

\bibitem[\protect\BCAY{Pal, Sekh, Kar,\ \BBA\ Prasad}{Pal
  et~al.}{2020}]{pal2020neural}
Pal, R., Sekh, A.~A., Kar, S., \BBA\ Prasad, D.~K. \BBOP2020\BBCP.
\newblock \BBOQ Neural network based country wise risk prediction of
  {COVID}-19\BBCQ\
\newblock {\Bem arXiv preprint arXiv:2004.00959}.

\bibitem[\protect\BCAY{Pandey, Gautam, Bhagat,\ \BBA\ Sethi}{Pandey
  et~al.}{2020}]{pandey2020wash}
Pandey, R., Gautam, V., Bhagat, K., \BBA\ Sethi, T. \BBOP2020\BBCP.
\newblock \BBOQ A machine learning application for raising {WASH} awareness in
  the times of {COVID}-19 pandemic\BBCQ\
\newblock {\Bem arXiv preprint arXiv:2003.07074}.

\bibitem[\protect\BCAY{{Penn Medicine}}{{Penn Medicine}}{2020}]{chime}
{Penn Medicine} \BBOP2020\BBCP.
\newblock \BBOQ {COVID}-19 hospital impact model for epidemics ({CHIME})\BBCQ\
\newblock \urllink{https://penn-chime.phl.io/}{https://penn-chime.phl.io/}.

\bibitem[\protect\BCAY{Pepe, Bajardi, Gauvin, Privitera, Lake, Cattuto,\ \BBA\
  Tizzoni}{Pepe et~al.}{2020}]{pepe2020covid}
Pepe, E., Bajardi, P., Gauvin, L., Privitera, F., Lake, B., Cattuto, C., \BBA\
  Tizzoni, M. \BBOP2020\BBCP.
\newblock \BBOQ {COVID}-19 outbreak response, a dataset to assess mobility
  changes in {Italy} following national lockdown\BBCQ\
\newblock {\Bem Scientific Data}, {\Bem 7\/}(1), 1--7.

\bibitem[\protect\BCAY{Qi, Jiang, Yu, Shao, Zhang, Yue, Ma, Wang, Liu, Meng,
  et~al.}{Qi et~al.}{2020}]{qi2020machine}
Qi, X., Jiang, Z., Yu, Q., Shao, C., Zhang, H., Yue, H., Ma, B., Wang, Y., Liu,
  C., Meng, X., et~al. \BBOP2020\BBCP.
\newblock \BBOQ Machine learning-based {CT} radiomics model for predicting
  hospital stay in patients with pneumonia associated with {SARS}-{CoV}-2
  infection: A multicenter study\BBCQ\
\newblock {\Bem medRxiv preprint medRxiv:2020.02.29.20029603}.

\bibitem[\protect\BCAY{Qing, Lee, De~Raeymaecker, Tame, Zhang, De~Maeyer,\
  \BBA\ Voet}{Qing et~al.}{2014}]{qing_pharmacophore_2014}
Qing, X., Lee, X.~Y., De~Raeymaecker, J., Tame, J.~R., Zhang, K.~Y., De~Maeyer,
  M., \BBA\ Voet, A. \BBOP2014\BBCP.
\newblock \BBOQ Pharmacophore modeling: Advances, limitations, and current
  utility in drug discovery\BBCQ\
\newblock {\Bem Journal of Receptor, Ligand and Channel Research}, {\Bem 7},
  81--92.

\bibitem[\protect\BCAY{Radin, Wineinger, Topol,\ \BBA\ Steinhubl}{Radin
  et~al.}{2020}]{radin2020harnessing}
Radin, J.~M., Wineinger, N.~E., Topol, E.~J., \BBA\ Steinhubl, S.~R.
  \BBOP2020\BBCP.
\newblock \BBOQ Harnessing wearable device data to improve state-level
  real-time surveillance of influenza-like illness in the {USA}: {A}
  population-based study\BBCQ\
\newblock {\Bem The Lancet Digital Health}, {\Bem 2}, e85--–93.

\bibitem[\protect\BCAY{Raghu\ \BBA\ Schmidt}{Raghu\ \BBA\
  Schmidt}{2020}]{raghu2020survey}
Raghu, M.\BBACOMMA\  \BBA\ Schmidt, E. \BBOP2020\BBCP.
\newblock \BBOQ A survey of deep learning for scientific discovery\BBCQ\
\newblock {\Bem arXiv preprint arXiv:2003.11755}.

\bibitem[\protect\BCAY{Randhawa, Soltysiak, El~Roz, de~Souza, Hill,\ \BBA\
  Kari}{Randhawa et~al.}{2020}]{randhawa_machine_2020}
Randhawa, G.~S., Soltysiak, M.~P., El~Roz, H., de~Souza, C.~P., Hill, K.~A.,
  \BBA\ Kari, L. \BBOP2020\BBCP.
\newblock \BBOQ Machine learning using intrinsic genomic signatures for rapid
  classification of novel pathogens: {COVID}-19 case study\BBCQ\
\newblock {\Bem {PLOS One}}, {\Bem 15\/}(4), 1--24.

\bibitem[\protect\BCAY{Rao\ \BBA\ Vazquez}{Rao\ \BBA\
  Vazquez}{2020}]{rao2020identification}
Rao, A. S.~S.\BBACOMMA\  \BBA\ Vazquez, J.~A. \BBOP2020\BBCP.
\newblock \BBOQ Identification of {COVID}-19 can be quicker through artificial
  intelligence framework using a mobile phone-based survey in the populations
  when cities/towns are under quarantine\BBCQ\
\newblock {\Bem Infection Control \& Hospital Epidemiology}, {\Bem 41\/}(7),
  826--830.

\bibitem[\protect\BCAY{Richardson, Griffin, Tucker, Smith, Oechsle, Phelan,\
  \BBA\ Stebbing}{Richardson et~al.}{2020}]{richardson_baricitinib_2020}
Richardson, P., Griffin, I., Tucker, C., Smith, D., Oechsle, O., Phelan, A.,
  \BBA\ Stebbing, J. \BBOP2020\BBCP.
\newblock \BBOQ Baricitinib as potential treatment for 2019-{{nCoV}} acute
  respiratory disease\BBCQ\
\newblock {\Bem The Lancet}, {\Bem 395\/}(10223), e30--e31.

\bibitem[\protect\BCAY{Ronneberger, Fischer,\ \BBA\ Brox}{Ronneberger
  et~al.}{2015}]{ronneberger2015}
Ronneberger, O., Fischer, P., \BBA\ Brox, T. \BBOP2015\BBCP.
\newblock \BBOQ U-net: {C}onvolutional networks for biomedical image
  segmentation\BBCQ\
\newblock In {\Bem International conference on medical image computing and
  computer-assisted intervention}, \BPGS\ 234--241. Springer.

\bibitem[\protect\BCAY{Ronsivalle, Foresti,\ \BBA\ Poledda}{Ronsivalle
  et~al.}{2020}]{ronsivalle2020prototype}
Ronsivalle, G.~B., Foresti, L., \BBA\ Poledda, G. \BBOP2020\BBCP.
\newblock \BBOQ A prototype model of georeferencing the inherent risk of
  contagion from {COVID}-19\BBCQ\
\newblock {\Bem ResearchGate preprint}.

\bibitem[\protect\BCAY{Roy\ \BBA\ Karmakar}{Roy\ \BBA\
  Karmakar}{2020}]{roy2020bayesian}
Roy, A.\BBACOMMA\  \BBA\ Karmakar, S. \BBOP2020\BBCP.
\newblock \BBOQ Bayesian semiparametric time varying model for count data to
  study the spread of the {COVID}-19 cases\BBCQ\
\newblock {\Bem arXiv preprint arXiv:2004.02281}.

\bibitem[\protect\BCAY{Safegraph}{Safegraph}{2020}]{foottraffic}
Safegraph \BBOP2020\BBCP.
\newblock \BBOQ The impact of {Coronavirus} ({COVID}-19) on foot traffic\BBCQ\
\newblock
  \urllink{https://www.safegraph.com/dashboard/covid19-commerce-patterns?is=5e7a3815f20d617a17a33173}{https://www.safegraph.com/dashboard/covid19-commerce-patterns?is5e7a3815f20d6
  17a17a33173}.

\bibitem[\protect\BCAY{Schild, Ling, Blackburn, Stringhini, Zhang,\ \BBA\
  Zannettou}{Schild et~al.}{2020}]{schild2020go}
Schild, L., Ling, C., Blackburn, J., Stringhini, G., Zhang, Y., \BBA\
  Zannettou, S. \BBOP2020\BBCP.
\newblock \BBOQ ``go eat a bat, {C}hang!”: An early look on the emergence of
  {S}inophobic behavior on web communities in the face of {COVID}-19\BBCQ\
\newblock {\Bem arXiv preprint arXiv:2004.04046}.

\bibitem[\protect\BCAY{Segler, Preuss,\ \BBA\ Waller}{Segler
  et~al.}{2018}]{segler_planning_2018}
Segler, M.~H., Preuss, M., \BBA\ Waller, M.~P. \BBOP2018\BBCP.
\newblock \BBOQ Planning chemical syntheses with deep neural networks and
  symbolic {{AI}}\BBCQ\
\newblock {\Bem Nature}, {\Bem 555\/}(7698), 604--610.

\bibitem[\protect\BCAY{Senior, Jumper, Hassabis,\ \BBA\ Kohli}{Senior
  et~al.}{2020a}]{senior_alphafold_2020}
Senior, A., Jumper, J., Hassabis, D., \BBA\ Kohli, P. \BBOP2020a\BBCP.
\newblock \BBOQ {{AlphaFold}}: {{Using AI}} for scientific discovery\BBCQ\
\newblock
  \urllink{https://deepmind.com/blog/article/AlphaFold-Using-AI-for-scientific-discovery}{https://deepmind.com/blog/article/AlphaFold-Using-AI-for-scientific-discovery}.

\bibitem[\protect\BCAY{Senior, Evans, Jumper, Kirkpatrick, Sifre, Green, Qin,
  {\v Z}{\'i}dek, Nelson, Bridgland, Penedones, Petersen, Simonyan, Crossan,
  Kohli, Jones, Silver, Kavukcuoglu,\ \BBA\ Hassabis}{Senior
  et~al.}{2020b}]{senior_improved_2020}
Senior, A.~W., Evans, R., Jumper, J., Kirkpatrick, J., Sifre, L., Green, T.,
  Qin, C., {\v Z}{\'i}dek, A., Nelson, A. W.~R., Bridgland, A., Penedones, H.,
  Petersen, S., Simonyan, K., Crossan, S., Kohli, P., Jones, D.~T., Silver, D.,
  Kavukcuoglu, K., \BBA\ Hassabis, D. \BBOP2020b\BBCP.
\newblock \BBOQ Improved protein structure prediction using potentials from
  deep learning\BBCQ\
\newblock {\Bem Nature}, {\Bem 577\/}(7792), 706--710.

\bibitem[\protect\BCAY{Senior, Evans, Jumper, Kirkpatrick, Sifre, Green, Qin,
  {\v Z}{\'i}dek, Nelson,\ \BBA\ Bridgland}{Senior
  et~al.}{2019}]{senior_protein_2019}
Senior, A.~W., Evans, R., Jumper, J., Kirkpatrick, J., Sifre, L., Green, T.,
  Qin, C., {\v Z}{\'i}dek, A., Nelson, A.~W., \BBA\ Bridgland, A.
  \BBOP2019\BBCP.
\newblock \BBOQ Protein structure prediction using multiple deep neural
  networks in the 13th {{Critical Assessment}} of {{Protein Structure
  Prediction}} ({{CASP13}})\BBCQ\
\newblock {\Bem Proteins: Structure, Function, and Bioinformatics}, {\Bem
  87\/}(12), 1141--1148.

\bibitem[\protect\BCAY{Shan, Gao, Wang, Shi, Shi, Han, Xue,\ \BBA\ Shi}{Shan
  et~al.}{2020}]{shan2020lung}
Shan, F., Gao, Y., Wang, J., Shi, W., Shi, N., Han, M., Xue, Z., \BBA\ Shi, Y.
  \BBOP2020\BBCP.
\newblock \BBOQ Lung infection quantification of {COVID}-19 in {CT} images with
  deep learning\BBCQ\
\newblock {\Bem arXiv preprint arXiv:2003.04655}.

\bibitem[\protect\BCAY{Shi, Peng, Liu, Cheng, Lu, Yang, Zhang, Li, Wang, Zhang,
  Gao, Shi, Zhang,\ \BBA\ Shan}{Shi et~al.}{2020}]{shi2020deep}
Shi, W., Peng, X., Liu, T., Cheng, Z., Lu, H., Yang, S., Zhang, J., Li, F.,
  Wang, M., Zhang, X., Gao, Y., Shi, Y., Zhang, Z., \BBA\ Shan, F.
  \BBOP2020\BBCP.
\newblock \BBOQ Deep learning-based quantitative computed tomography model in
  predicting the severity of {COVID}-19: A retrospective study in 196
  patients\BBCQ\
\newblock {\Bem The Lancet preprint}.

\bibitem[\protect\BCAY{Singh, Valley, Tang, Li, Kamran, Sjoding, Wiens, Otles,
  Donnelly, Wei, McBride, Cao, Penoza, Ayanian,\ \BBA\ Nallamothu}{Singh
  et~al.}{2020}]{singh2020validating}
Singh, K., Valley, T.~S., Tang, S., Li, B.~Y., Kamran, F., Sjoding, M.~W.,
  Wiens, J., Otles, E., Donnelly, J.~P., Wei, M.~Y., McBride, J.~P., Cao, J.,
  Penoza, C., Ayanian, J.~Z., \BBA\ Nallamothu, B.~K. \BBOP2020\BBCP.
\newblock \BBOQ Validating a widely implemented deterioration index model among
  hospitalized {COVID}-19 patients\BBCQ\
\newblock {\Bem medRxiv preprint medRxiv:2020.04.24.20079012}.

\bibitem[\protect\BCAY{Singha, Bansala, Bodea, Budakb, Chic, Kawintiranona,
  Paddena, Vanarsdalla, Vragad,\ \BBA\ Wanga}{Singha
  et~al.}{2020}]{singh2020first}
Singha, L., Bansala, S., Bodea, L., Budakb, C., Chic, G., Kawintiranona, K.,
  Paddena, C., Vanarsdalla, R., Vragad, E., \BBA\ Wanga, Y. \BBOP2020\BBCP.
\newblock \BBOQ A first look at {COVID}-19 information and misinformation
  sharing on {Twitter}\BBCQ\
\newblock {\Bem arXiv preprint arXiv:2003.13907}.

\bibitem[\protect\BCAY{Song, Zheng, Li, Zhang, Zhang, Huang, Chen, Zhao, Jie,
  Wang, et~al.}{Song et~al.}{2020}]{song2020deep}
Song, Y., Zheng, S., Li, L., Zhang, X., Zhang, X., Huang, Z., Chen, J., Zhao,
  H., Jie, Y., Wang, R., et~al. \BBOP2020\BBCP.
\newblock \BBOQ Deep learning enables accurate diagnosis of novel coronavirus
  ({COVID}-19) with {CT} images\BBCQ\
\newblock {\Bem medRxiv preprint medRxiv:2020.02.23.20026930}.

\bibitem[\protect\BCAY{Sterling\ \BBA\ Irwin}{Sterling\ \BBA\
  Irwin}{2015}]{sterling2015zinc}
Sterling, T.\BBACOMMA\  \BBA\ Irwin, J.~J. \BBOP2015\BBCP.
\newblock \BBOQ {ZINC} 15 -- {L}igand discovery for everyone\BBCQ\
\newblock {\Bem Journal of Chemical Information and Modeling}, {\Bem 55\/}(11),
  2324--2337.

\bibitem[\protect\BCAY{Su, Luo, Castro, Moos, McFarland, Emanuele, Kassel,\
  \BBA\ Zhuangfang}{Su et~al.}{2020}]{covidcare}
Su, A., Luo, D., Castro, H., Moos, L., McFarland, M., Emanuele, R., Kassel, S.,
  \BBA\ Zhuangfang, N.~Y. \BBOP2020\BBCP.
\newblock \BBOQ {COVID}-19 healthcare system capacity\BBCQ\
\newblock
  \urllink{https://github.com/covidcaremap/covid19-healthsystemcapacity}{https://github.com/covidcaremap/covid19-healthsystemcapacity}.

\bibitem[\protect\BCAY{Szegedy, Liu, Jia, Sermanet, Reed, Anguelov, Erhan,
  Vanhoucke,\ \BBA\ Rabinovich}{Szegedy et~al.}{2015}]{szegedy2015going}
Szegedy, C., Liu, W., Jia, Y., Sermanet, P., Reed, S., Anguelov, D., Erhan, D.,
  Vanhoucke, V., \BBA\ Rabinovich, A. \BBOP2015\BBCP.
\newblock \BBOQ Going deeper with convolutions\BBCQ\
\newblock In {\Bem Proceedings of the IEEE conference on computer vision and
  pattern recognition}, \BPGS\ 1--9.

\bibitem[\protect\BCAY{Tang, He, Liu, Fang, Wu,\ \BBA\ Xu}{Tang
  et~al.}{2020a}]{tang_ai-aided_2020}
Tang, B., He, F., Liu, D., Fang, M., Wu, Z., \BBA\ Xu, D. \BBOP2020a\BBCP.
\newblock \BBOQ {{AI}}-aided design of novel targeted covalent inhibitors
  against {{SARS}}-{{CoV}}-2\BBCQ\
\newblock {\Bem bioRxiv preprint bioRxiv:2020.03.03.972133}.

\bibitem[\protect\BCAY{Tang, Zhao, Xie, Zhong, Shi, Liu,\ \BBA\ Shen}{Tang
  et~al.}{2020b}]{tang2020severity}
Tang, Z., Zhao, W., Xie, X., Zhong, Z., Shi, F., Liu, J., \BBA\ Shen, D.
  \BBOP2020b\BBCP.
\newblock \BBOQ Severity assessment of coronavirus disease 2019 ({COVID}-19)
  using quantitative features from chest {CT} images\BBCQ\
\newblock {\Bem arXiv preprint arXiv:2003.11988}.

\bibitem[\protect\BCAY{Ton, Gentile, Hsing, Ban,\ \BBA\ Cherkasov}{Ton
  et~al.}{2020}]{ton_rapid_2020}
Ton, A.-T., Gentile, F., Hsing, M., Ban, F., \BBA\ Cherkasov, A.
  \BBOP2020\BBCP.
\newblock \BBOQ Rapid identification of potential inhibitors of
  {{SARS}}-{{CoV}}-2 main protease by deep docking of 1.3 billion
  compounds\BBCQ\
\newblock {\Bem Molecular Informatics}, {\Bem 39}, 1--8.

\bibitem[\protect\BCAY{{United Nations}}{{United Nations}}{2020}]{UNhatespeech}
{United Nations} \BBOP2020\BBCP.
\newblock \BBOQ United {N}ations guidance note on addressing and countering
  {COVID}-19 related hate speech\BBCQ\
\newblock
  \urllink{https://www.un.org/en/genocideprevention/documents/Guidance\%20on\%20COVID-19\%20related\%20Hate\%20Speech.pdf}{https://www.un.org/en/genocideprevention/
  documents/Guidance\%20on\%20COVID-19\%20related\%20Hate\%20Speech.pdf}.

\bibitem[\protect\BCAY{Velásquez, Leahy, Restrepo, Lupu, Sear, Gabriel, Jha,\
  \BBA\ Johnson}{Velásquez et~al.}{2020}]{velasquez2020hate}
Velásquez, N., Leahy, R., Restrepo, N.~J., Lupu, Y., Sear, R., Gabriel, N.,
  Jha, O., \BBA\ Johnson, N. \BBOP2020\BBCP.
\newblock \BBOQ Hate multiverse spreads malicious {COVID}-19 content online
  beyond individual platform control\BBCQ\
\newblock {\Bem arXiv preprint arXiv:2004.00673}.

\bibitem[\protect\BCAY{Wang\ \BBA\ Wong}{Wang\ \BBA\
  Wong}{2020}]{wang2020covid}
Wang, L.\BBACOMMA\  \BBA\ Wong, A. \BBOP2020\BBCP.
\newblock \BBOQ {COVID-Net}: A tailored deep convolutional neural network
  design for detection of {COVID}-19 cases from chest radiography images\BBCQ\
\newblock {\Bem arXiv preprint arXiv:2003.09871}.

\bibitem[\protect\BCAY{Wang, Lo, Chandrasekhar, Reas, Yang, Eide, Funk, Kinney,
  Liu,\ \BBA\ Merrill}{Wang et~al.}{2020}]{wang_cord-19_2020}
Wang, L.~L., Lo, K., Chandrasekhar, Y., Reas, R., Yang, J., Eide, D., Funk, K.,
  Kinney, R., Liu, Z., \BBA\ Merrill, W. \BBOP2020\BBCP.
\newblock \BBOQ {{CORD}}-19: {T}he {COVID}-19 open research dataset\BBCQ\
\newblock {\Bem arXiv preprint arXiv:2004.10706}.

\bibitem[\protect\BCAY{Wang, Fang, Lu,\ \BBA\ Wang}{Wang
  et~al.}{2004}]{wang2004pdbbind}
Wang, R., Fang, X., Lu, Y., \BBA\ Wang, S. \BBOP2004\BBCP.
\newblock \BBOQ The {PDBbind} database: Collection of binding affinities for
  protein-ligand complexes with known three-dimensional structures\BBCQ\
\newblock {\Bem Journal of Medicinal Chemistry}, {\Bem 47\/}(12), 2977--2980.

\bibitem[\protect\BCAY{Wang, Kang, Ma, Zeng, Xiao, Guo, Cai, Yang, Li, Meng,
  et~al.}{Wang et~al.}{2020a}]{wang2020deep}
Wang, S., Kang, B., Ma, J., Zeng, X., Xiao, M., Guo, J., Cai, M., Yang, J., Li,
  Y., Meng, X., et~al. \BBOP2020a\BBCP.
\newblock \BBOQ A deep learning algorithm using {CT} images to screen for
  corona virus disease ({COVID}-19)\BBCQ\
\newblock {\Bem medRxiv preprint medRxiv:2020.02.14.20023028}.

\bibitem[\protect\BCAY{Wang, Hu, Li, Zhang, Zhai,\ \BBA\ Yao}{Wang
  et~al.}{2020b}]{wang2020abnormal}
Wang, Y., Hu, M., Li, Q., Zhang, X.-P., Zhai, G., \BBA\ Yao, N.
  \BBOP2020b\BBCP.
\newblock \BBOQ Abnormal respiratory patterns classifier may contribute to
  large-scale screening of people infected with {COVID}-19 in an accurate and
  unobtrusive manner\BBCQ\
\newblock {\Bem arXiv preprint arXiv:2002.05534}.

\bibitem[\protect\BCAY{Weinstock, Echenique,\ \BBA\ Russell}{Weinstock
  et~al.}{2020}]{weinstock2020}
Weinstock, M., Echenique, A., \BBA\ Russell, J. e.~a. \BBOP2020\BBCP.
\newblock \BBOQ Chest {X}-ray findings in 636 ambulatory patients with
  {COVID}-19 presenting to an urgent care center: {A} normal chest {X}-ray is
  no guarantee\BBCQ\
\newblock {\Bem The Journal of Urgent Care Medicine}, {\Bem 14\/}(7), 13--18.

\bibitem[\protect\BCAY{WHO}{WHO}{2020a}]{who2020sitrep}
WHO \BBOP2020a\BBCP.
\newblock \BBOQ Coronavirus disease (covid-2019) situation reports\BBCQ\
\newblock
  \urllink{https://www.who.int/emergencies/diseases/novel-coronavirus-2019}{https://www.who.int/
  emergencies/diseases/novel-coronavirus-2019}.

\bibitem[\protect\BCAY{WHO}{WHO}{2020b}]{who2020vaccines}
WHO \BBOP2020b\BBCP.
\newblock \BBOQ Draft landscape of {COVID}-19 candidate vaccines\BBCQ\
\newblock
  \urllink{https://www.who.int/who-documents-detail/draft-landscape-of-covid-19-candidate-vaccines}{https://www.who.int/
  who-documents-detail/draft-landscape-of-covid-19-candidate-vaccines}.

\bibitem[\protect\BCAY{WHO}{WHO}{2020c}]{who2020infordmic}
WHO \BBOP2020c\BBCP.
\newblock \BBOQ Infodemic management - {I}nfodemiology\BBCQ\
\newblock
  \urllink{https://www.who.int/teams/risk-communication/infodemic-management}{https://www.who.int/teams/risk-communication/infodemic-management}.

\bibitem[\protect\BCAY{WHO}{WHO}{2020d}]{who2020chatbot}
WHO \BBOP2020d\BBCP.
\newblock \BBOQ {WHO} and {R}akuten {V}iber fight {COVID}-19 misinformation
  with interactive chatbot\BBCQ\
\newblock
  \urllink{https://www.who.int/news-room/feature-stories/detail/who-and-rakuten-viber-fight-covid-19-misinformation-with-interactive-chatbot}{https://www.who.int/news-room/feature-stories/detail/who-and-rakuten-viber-fight-covid-19-misinformation-with-interactive-chatbot}.

\bibitem[\protect\BCAY{WHO}{WHO}{2020e}]{who2020database}
WHO \BBOP2020e\BBCP.
\newblock \BBOQ The {WHO} research database on {COVID}-19\BBCQ\
\newblock
  \urllink{https://www.who.int/emergencies/diseases/novel-coronavirus-2019/global-research-on-novel-coronavirus-2019-ncov}{https://www.who.int/
  emergencies/diseases/novel-coronavirus-2019/global-research-on-novel-coronavirus-2019-ncov}.

\bibitem[\protect\BCAY{Wishart, Feunang, Guo, Lo, Marcu, Grant, Sajed, Johnson,
  Li, Sayeeda, et~al.}{Wishart et~al.}{2018}]{wishart2018drugbank}
Wishart, D.~S., Feunang, Y.~D., Guo, A.~C., Lo, E.~J., Marcu, A., Grant, J.~R.,
  Sajed, T., Johnson, D., Li, C., Sayeeda, Z., et~al. \BBOP2018\BBCP.
\newblock \BBOQ {DrugBank} 5.0: A major update to the {DrugBank} database for
  2018\BBCQ\
\newblock {\Bem Nucleic Acids Research}, {\Bem 46\/}(D1), D1074--D1082.

\bibitem[\protect\BCAY{Wynants, Van~Calster, Bonten, Collins, Debray, De~Vos,
  Haller, Heinze, Moons, Riley, et~al.}{Wynants
  et~al.}{2020}]{wynants2020prediction}
Wynants, L., Van~Calster, B., Bonten, M.~M., Collins, G.~S., Debray, T.~P.,
  De~Vos, M., Haller, M.~C., Heinze, G., Moons, K.~G., Riley, R.~D., et~al.
  \BBOP2020\BBCP.
\newblock \BBOQ Prediction models for diagnosis and prognosis of {COVID-19}
  infection: {S}ystematic review and critical appraisal\BBCQ\
\newblock {\Bem BMJ}, {\Bem 369}.

\bibitem[\protect\BCAY{Xu, Gutierrez, Mekaru, Sewalk, Goodwin, Loskill, Cohn,
  Hswen, Hill, Cobo, et~al.}{Xu et~al.}{2020a}]{xu2020epidemiological}
Xu, B., Gutierrez, B., Mekaru, S., Sewalk, K., Goodwin, L., Loskill, A., Cohn,
  E.~L., Hswen, Y., Hill, S.~C., Cobo, M.~M., et~al. \BBOP2020a\BBCP.
\newblock \BBOQ Epidemiological data from the {COVID}-19 outbreak, real-time
  case information\BBCQ\
\newblock {\Bem Scientific Data}, {\Bem 7\/}(1), 1--6.

\bibitem[\protect\BCAY{Xu, Jiang, Ma, Du, Li, Lv, Yu, Ni, Chen, Su, et~al.}{Xu
  et~al.}{2020b}]{xu2020deep}
Xu, X., Jiang, X., Ma, C., Du, P., Li, X., Lv, S., Yu, L., Ni, Q., Chen, Y.,
  Su, J., et~al. \BBOP2020b\BBCP.
\newblock \BBOQ A deep learning system to screen novel coronavirus disease 2019
  pneumonia\BBCQ\
\newblock {\Bem Engineering}.

\bibitem[\protect\BCAY{Yan, Zhang, Xiao, Wang, Sun, Liang, Li, Zhang, Guo,
  Xiao, et~al.}{Yan et~al.}{2020}]{yan2020prediction}
Yan, L., Zhang, H.-T., Xiao, Y., Wang, M., Sun, C., Liang, J., Li, S., Zhang,
  M., Guo, Y., Xiao, Y., et~al. \BBOP2020\BBCP.
\newblock \BBOQ Prediction of criticality in patients with severe {COVID}-19
  infection using three clinical features: a machine learning-based prognostic
  model with clinical data in {W}uhan\BBCQ\
\newblock {\Bem medRxiv preprint medRxiv:L2020.02.27.20028027}.

\bibitem[\protect\BCAY{Yang, Anishchenko, Park, Peng, Ovchinnikov,\ \BBA\
  Baker}{Yang et~al.}{2020}]{yang_improved_2020}
Yang, J., Anishchenko, I., Park, H., Peng, Z., Ovchinnikov, S., \BBA\ Baker, D.
  \BBOP2020\BBCP.
\newblock \BBOQ Improved protein structure prediction using predicted
  interresidue orientations\BBCQ\
\newblock {\Bem Proceedings of the National Academy of Sciences}, {\Bem
  117\/}(3), 1496--1503.

\bibitem[\protect\BCAY{Yang, Santillana,\ \BBA\ Kou}{Yang
  et~al.}{2015}]{yang2015argo}
Yang, S., Santillana, M., \BBA\ Kou, S.~C. \BBOP2015\BBCP.
\newblock \BBOQ Accurate estimation of influenza epidemics using {G}oogle
  search data via {{ARGO}}\BBCQ\
\newblock {\Bem Proceedings of the National Academy of Sciences}, {\Bem
  112\/}(47), 14473--14478.

\bibitem[\protect\BCAY{Yang}{Yang}{2012}]{yang2012flower}
Yang, X. \BBOP2012\BBCP.
\newblock \BBOQ Flower pollination algorithm for global optimization\BBCQ\
\newblock {\Bem International Conference on Unconventional Computing and
  Natural Computation}, 240--249.

\bibitem[\protect\BCAY{{Ye}, {Hou}, {Fan}, {Zhang}, {Qian}, {Sun}, {Peng},
  {Ju}, {Song},\ \BBA\ {Loparo}}{{Ye} et~al.}{2020}]{ye2020alpha}
{Ye}, Y., {Hou}, S., {Fan}, Y., {Zhang}, Y., {Qian}, Y., {Sun}, S., {Peng}, Q.,
  {Ju}, M., {Song}, W., \BBA\ {Loparo}, K. \BBOP2020\BBCP.
\newblock \BBOQ $alpha$-{S}atellite: {A}n {AI}-driven system and benchmark
  datasets for dynamic {COVID}-19 risk assessment in the {U}nited
  {S}tates\BBCQ\
\newblock {\Bem IEEE Journal of Biomedical and Health Informatics}, 1--1.

\bibitem[\protect\BCAY{Yu\ \BBA\ Koltun}{Yu\ \BBA\
  Koltun}{2016}]{yu_multi-scale_2016}
Yu, F.\BBACOMMA\  \BBA\ Koltun, V. \BBOP2016\BBCP.
\newblock \BBOQ Multi-scale context aggregation by dilated convolutions\BBCQ\
\newblock In {\Bem Proceedings of the 2016 International Conference on Learning
  Representations}.

\bibitem[\protect\BCAY{Yu}{Yu}{2020}]{yu2020open}
Yu, J. \BBOP2020\BBCP.
\newblock \BBOQ Open access institutional and news media tweet dataset for
  {COVID}-19 social science research\BBCQ\
\newblock {\Bem arXiv preprint arXiv:2004.01791}.

\bibitem[\protect\BCAY{Zarocostas}{Zarocostas}{2020}]{zarocostas2020lancet}
Zarocostas, J. \BBOP2020\BBCP.
\newblock \BBOQ How to fight an infodemic\BBCQ\
\newblock {\Bem The Lancet}, {\Bem 395\/}(10225), 676.

\bibitem[\protect\BCAY{Zhang, Saravanan, Yang, Hossain, Li, Ren,\ \BBA\
  Wei}{Zhang et~al.}{2020}]{zhang_deep_2020}
Zhang, H., Saravanan, K.~M., Yang, Y., Hossain, M.~T., Li, J., Ren, X., \BBA\
  Wei, Y. \BBOP2020\BBCP.
\newblock \BBOQ Deep learning based drug screening for novel coronavirus
  2019-{{nCov}}\BBCQ\
\newblock {\Bem Interdisciplinary Sciences: Computational Life Sciences}.

\bibitem[\protect\BCAY{Zhao, Song, Chen, Zou, Ma, Ma, Li, Hao, Li, Tian,
  et~al.}{Zhao et~al.}{2020}]{zhao_2020_novel}
Zhao, W.-M., Song, S.-H., Chen, M.-L., Zou, D., Ma, L.-N., Ma, Y.-K., Li,
  R.-J., Hao, L.-L., Li, C.-P., Tian, D.-M., et~al. \BBOP2020\BBCP.
\newblock \BBOQ The 2019 novel coronavirus resource\BBCQ\
\newblock {\Bem Yi Chuan = Hereditas}, {\Bem 42\/}(2), 212--221.

\bibitem[\protect\BCAY{Zhavoronkov}{Zhavoronkov}{2018}]{zhavoronkov_artificial_2018}
Zhavoronkov, A. \BBOP2018\BBCP.
\newblock \BBOQ Artificial intelligence for drug discovery, biomarker
  development, and generation of novel chemistry\BBCQ\
\newblock {\Bem Molecular Pharmaceutics}, {\Bem 15\/}(10), 4311--4313.

\bibitem[\protect\BCAY{Zhavoronkov, Aladinskiy, Zhebrak, Zagribelnyy,
  Terentiev, Bezrukov, Polykovskiy, Shayakhmetov, Filimonov, Orekhov,
  et~al.}{Zhavoronkov et~al.}{2020}]{zhavoronkov2020potential}
Zhavoronkov, A., Aladinskiy, V., Zhebrak, A., Zagribelnyy, B., Terentiev, V.,
  Bezrukov, D.~S., Polykovskiy, D., Shayakhmetov, R., Filimonov, A., Orekhov,
  P., et~al. \BBOP2020\BBCP.
\newblock \BBOQ Potential {COVID}-2019 {3C}-like protease inhibitors designed
  using generative deep learning approaches\BBCQ\
\newblock {\Bem chemRxiv preprint chemrxiv:11829102.v2}.

\bibitem[\protect\BCAY{Zheng, Li, Zhang, Pearce, Mortuza,\ \BBA\ Zhang}{Zheng
  et~al.}{2019}]{zheng_deep-learning_2019}
Zheng, W., Li, Y., Zhang, C., Pearce, R., Mortuza, S.~M., \BBA\ Zhang, Y.
  \BBOP2019\BBCP.
\newblock \BBOQ Deep-learning contact-map guided protein structure prediction
  in {{CASP13}}\BBCQ\
\newblock {\Bem Proteins: Structure, Function, and Bioinformatics}, {\Bem
  87\/}(12), 1149--1164.

\bibitem[\protect\BCAY{Zhou, Rahman~Siddiquee, Tajbakhsh,\ \BBA\ Liang}{Zhou
  et~al.}{2018}]{zhou2018unet}
Zhou, Z., Rahman~Siddiquee, M.~M., Tajbakhsh, N., \BBA\ Liang, J.
  \BBOP2018\BBCP.
\newblock \BBOQ Unet++: A nested {U}-net architecture for medical image
  segmentation\BBCQ\
\newblock In {\Bem Deep Learning in Medical Image Analysis and Multimodal
  Learning for Clinical Decision Support. DLMIA 2018, ML-CDS 2018. Lecture
  Notes in Computer Science}, \lowercase{\BVOL}\ 11045. Springer.

\bibitem[\protect\BCAY{Zou\ \BBA\ Hastie}{Zou\ \BBA\
  Hastie}{2005}]{zoue2005elasticnet}
Zou, H.\BBACOMMA\  \BBA\ Hastie, T. \BBOP2005\BBCP.
\newblock \BBOQ Regularization and variable selection via the elastic net\BBCQ\
\newblock {\Bem Journal of the Royal Statistical Society: Series B (Statistical
  Methodology)}, {\Bem 67\/}(2), 301--320.

\end{thebibliography}
\bibliographystyle{theapa}

\end{document}